\documentclass[preprint2, longabstract]{aastex}
\usepackage{color}
\usepackage{ulem}

\shorttitle{Host galaxy dependencies of SN Ia}
\shortauthors{Lampeitl et al.}

\begin{document}

\title{The Effect of Host Galaxies on Type Ia Supernovae in the SDSS-II Supernova Survey}

\author{Hubert~Lampeitl\altaffilmark{1}, 
            Mathew~Smith\altaffilmark{2,1}, 
            Robert~C.~Nichol\altaffilmark{1}, 
            Bruce~Bassett\altaffilmark{3,4},
            David~Cinabro\altaffilmark{5},
            Benjamin~Dilday\altaffilmark{6},
            Ryan~J.~Foley\altaffilmark{7,8},
            Joshua~A.~Frieman\altaffilmark{9,10},
            Peter~M.~Garnavich\altaffilmark{11},
            Ariel~Goobar\altaffilmark{12,13},
            Myungshin~Im\altaffilmark{14},
            Saurabh~W.~Jha\altaffilmark{6},
            John~Marriner\altaffilmark{10},
            Ramon~Miquel\altaffilmark{15,16},
            Jakob~Nordin\altaffilmark{12,13},
            Linda~\"{O}stman\altaffilmark{16},
            Adam~G.~Riess\altaffilmark{17,18}, 
            Masao~Sako\altaffilmark{19},
            Donald~P.~Schneider\altaffilmark{20},
            Jesper~Sollerman\altaffilmark{21, 22},
            Maximilian~Stritzinger\altaffilmark{22, 23}
            }
\affil{}
\affil{}
\affil{$^{1}$Institute of Cosmology and Gravitation, University of Portsmouth, Portsmouth, PO1 3FX, UK}
\affil{$^{2}$Astrophysics, Cosmology and Gravity Centre (ACGC), Department of Mathematics and Applied Mathematics, University of Cape Town, Rondebosch, 7701, SA}
\affil{$^{3}$South African Astronomical Observatory, P.O. Box 9, Observatory 7935, South Africa}
\affil{$^{4}$Dept. of Mathematics and Applied Mathematics, University of Cape Town, Rondebosch, 7701, SA}
\affil{$^{5}$Dept. of Physics \& Astronomy, Wayne State University, Detroit, MI 48202, USA}
\affil{$^{6}$Dept. of Physics \& Astronomy, Rutgers the State University of New Jersey, Piscataway, NJ 08854, USA}
\affil{$^{7}$Harvard-Smithsonian Center for Astrophysics, 60 Garden Street, Cambridge, MA 02138, USA}
\affil{$^{8}$Clay Fellow}
\affil{$^{9}$Dept. of Astronomy \& Astrophysics, University of Chicago, Chicago, IL 60637, USA} 
\affil{$^{10}$Fermilab, P.O. Box 500, Batavia, IL 60510, USA}
\affil{$^{11}$Dept. of Physics, University of Notre Dame, Notre Dame, IN 46556, USA}
\affil{$^{12}$The Oskar Klein Centre fro Cosmoparticle Physics, Department of Physics, Albanova Stockholm University, SE-106 91 Stockholm, Sweden}
\affil{$^{13}$Department of Physics, Stockholm University, Albanova University Center, S-106 91 Stockholm, Sweden}
\affil{$^{14}$CEOU\/Dept. of Physics \& Astronomy, Seoul National University, Seoul, Korea}
\affil{$^{15}$Instituci\'o Catalana de Recerca i Estudis Avan\c{c}ats, E-08010 Barcelona, Spain}
\affil{$^{16}$Institut de F\'isica d'Altes Energies, E-08193 Bellaterra (Barcelona), Spain}
\affil{$^{17}$Dept. of Physics \& Astronomy, Johns Hopkins University, Baltimore, MD 21218, USA}
\affil{$^{18}$Space Telescope Science Institute, Baltimore, MD 21218, USA}
\affil{$^{19}$Physics \& Astronomy, University of Pennsylvania, Philadelphia, PA 19104}
\affil{$^{20}$Dept. of Astronomy \& Astrophysics, The Pennsylvania State University, University Park, PA 16802, USA}
\affil{$^{21}$Dept. of Astronomy, Stockholm University, SE-106 91 Stockholm, Sweden}
\affil{$^{22}$Dark Cosmology Centre, Niels Bohr Institute, University of Copenhagen, Juliane Maries Vej 30, 2100 Copenhagen, Denmark}
\affil{$^{23}$Carnegie Observatories, Las Campanas Observatory, Casilla 601, La Serena, Chile}
\affil{}
\affil{}

\email{Hubert.Lampeitl@port.ac.uk}

\begin{abstract}
We present an analysis of the host galaxy dependencies of Type Ia Supernovae (SNe Ia) from the full three year sample of the SDSS-II Supernova Survey. We re-discover, to high significance, the strong correlation between host galaxy type and the width of the observed SN light curve, i.e., fainter, quickly declining SNe Ia favor passive host galaxies, while brighter, slowly declining Ia's favor star-forming galaxies. We also find evidence (at between 2 to $3\sigma$) that SNe Ia are $\simeq0.1\pm0.04$ magnitudes brighter in passive host galaxies, than in star--forming hosts, after the SN Ia light curves have been standardized using the light curve shape and color variations: This difference in brightness is present in both the SALT2 and MCLS2k2 light curve fitting methodologies. We see evidence for differences in the SN Ia color relationship between passive and star--forming host galaxies, e.g., for the MLCS2k2 technique, we see that SNe Ia in passive hosts favor a dust law of 
$R_V = 1.0\pm0.2$,
while SNe Ia in star-forming hosts require 
$R_V = 1.8^{+0.2}_{-0.4}$.
The significance of these trends depends on the range of SN colors considered. We demonstrate that these effects can be parameterized using the stellar mass of the host galaxy (with a confidence of $>4\sigma$) and including this extra parameter provides a better statistical fit to our data. Our results suggest that future cosmological analyses of SN Ia samples should include host galaxy information. 
\end{abstract}
 
\section{Introduction}
Over the last decade, Type Ia Supernovae (SNe Ia) have become important cosmological probes as they can be used to measure distances to high redshift ($z\lesssim1.5$). 
In recent years, numerous samples of SN Ia have been compiled, e.g., CSP \citep{2006PASP..118....2H}, SNLS \citep{2006A&A...447...31A}, ESSENCE \citep{2007ApJ...666..694W}, SDSS \citep{2008AJ....135..338F}, CfA \citep{2009ApJ...700.1097H}, and combined, we are approaching $\sim1000$ spectroscopically-confirmed SNe Ia available 
for cosmological analysis \citep{2010arXiv1004.1711A}. With such large samples, it is becoming increasingly important to understand the systematic uncertainties (photometric calibration, SN color variations, etc.) associated 
with using SNe Ia for cosmology, including any additional physical parameters that could reduce the intrinsic scatter of the population. 

One such parameter could be related to the environment of the supernova. First and foremost, one would expect differences in the colors of SNe Ia based on the different dust content of 
their hosts, i.e., potential variations in local circumstellar dust around the progenitor star \citep{wang2005, 2008ApJ...686L.103G} and/or differences in the global dust content of 
different galaxy types\footnote{Dust in our own Galaxy is usually corrected for using the dust maps of \citet{1998ApJ...500..525S}}.  Despite these concerns, most analyses account for dust 
during the fitting of the supernova light curves by assuming a single absorption law ($R_V$) for all SNe, which minimizes the scatter around the Hubble diagram \citep{1998A&A...331..815T}. 
This process, however, has led to a dust law that is significantly different from the canonical value for our Galaxy ($R_V \approx 3.1$), e.g., \citet{2007ApJ...664L..13C} find $R_V \approx 1$ 
for nearby SNe Ia, while  \citet{2009ApJS..185...32K} obtained a best fit of $R_V = 2.18 \pm 0.14{\rm (stat)} \pm {\rm 0.48(sys)}$ for the first year sample from the Sloan Digital Sky Survey II (SDSS-II) Supernova Survey. In a recent near-infrared study 
of nearby SNe,  \citet{2010AJ....139..120F} found $R_V \approx 1-2$ for their whole sample, but obtained $R_V \approx 3.2$ if they exclude their most reddened objects. 
Alternatively, in a study of 80 nearby SNe Ia, \citet{2008A&A...487...19N} found 
a value of $R_V=1.75\pm0.27$ for their whole sample and a lower value of $R_V\sim1$ if they restrict the sample to low reddening values.
These differences could suggest that the effects of dust may also be dependent on the particular line of sight \citep{2007ApJ...666..694W} or on the inclination of the host 
galaxies \citep{2010arXiv1001.1744M}.

Secondly, the details of the supernova progenitor system could systematically vary between the different galaxy types. Our present theoretical understanding of SNe Ia suggests they are the thermonuclear explosion of a carbon-oxygen white dwarf which has reached the Chandrasekhar limit \citep{1973ApJ...186.1007W, 2000ARA&A..38..191H}. The mechanism for how the progenitor system accretes 
mass could be different between galaxy types, either accretion from a nearby companion star (which could have different metallicities depending on the stellar populations in the host galaxy types) or the merger with another white dwarf \citep{2009ApJ...699.2026R}.

Therefore, there are clear reasons to search for correlations between the properties of Ia's and the properties of their host galaxies. For example, there is a well-established difference between the rates of Ia's in passive and star-forming galaxies, potentially indicating two different paths or timescales for Ia's 
\citep{1979AJ.....84..985O, 1990PASP..102.1318V, 2005A&A...433..807M, 2006ApJ...648..868S}. There have also been indications that the host galaxy type correlates with the observed residuals on the SN Hubble diagram, even after standardizing each SN Ia 
\citep{2003MNRAS.340.1057S, 2008ApJ...685..752G, 2009arXiv0912.0929K}. For example, 
\citet{sullivan2010}
recently reported that SNe Ia in massive host galaxies are 0.08 magnitudes brighter than those in lower mass hosts after correction for the light curve shape and color (at a statistical significance of 4$\sigma$). 
Such correlations would have important consequences for supernova surveys and could improve the use of SNe Ia as ``standard candles" \citep{2009ApJ...699L.139W, sullivan2010}.

We investigate the environmental dependencies of SNe Ia by studying the residuals on the Hubble diagram (around the best fit cosmology) as a function of host galaxy type. We further ask if there are differences in the assumed dust law depending on the type of the host galaxy. In Section 2, we outline the data used in this analysis, which is taken from the full SDSS-II Supernova Survey \citep{2008AJ....135..338F}. This sample of SNe Ia has several advantages for such environmental studies including high survey efficiency, multi-color ($ugriz$) photometry for all host galaxies and a significant cosmological volume, thus providing a fair sampling of the galaxy distribution. Also, the overall SN rate, as a function of galaxy type, has been measured using this data \citep{smithm}

In Section 2, and Appendix A, we outline the details of our analysis using two, well-known, public light curve fitting procedures; SALT2 \citep{2007A&A...466...11G} and MLCS2k2 \citep{2007ApJ...659..122J, 2009ApJS..185...32K}. We also describe our methodology for defining passive and star-forming host galaxies.  In Section 3, we present our main results, while in Section 4, we discuss these results in light of other work in the field. We conclude in Section 5.  

\section{Observations and Data Analysis}

\begin{deluxetable}{lcc|cc}
\tabletypesize{\scriptsize}
\tablecaption{Number of SNe in our sample. 
\label{tbl_sample}}
\tablewidth{0pt}
\tablehead{
\colhead{Selection} & \multicolumn{2}{c}{Spec Confirm} & \multicolumn{2}{|c}{Total$^a$}\\
\colhead{} & \colhead{SALT} & \colhead{MLCS} & \multicolumn{1}{|c}{SALT} & \colhead{MLCS} 
}

\startdata
All SNe                 & \multicolumn{2}{c}{258} & \multicolumn{2}{|c}{361} \\ 
\hline
After LC cut     & 192 & 187 & 253 & 256 \\
After LC fitter limits   & 185 & 161 & 234 & 214 \\
Valid host galaxy type & 127 & 104 & 162 & 135 \\
\hline
Passive         & 27 & 24 & 40  & 35 \\
Star-forming  & 100 & 80 & 122  & 100 \\ \hline
 \enddata
 \tablenotetext{a}{Total SNe used in our analysis including spectroscopically--confirmed and photometrically--classified SNe Ia}
\end{deluxetable}

\begin{deluxetable}{llrrrrrrrr}
\tabletypesize{\scriptsize}
\tablecaption{
Host properties used in the main SALT2 analysis. Hosts with negligible star-formation rates are indicated with $N/A$ and
members of the restricted sample with r.
This is a sample of the full version of the table published in electronic format and can be found at the end of the preprint.}
\tablewidth{0pt}
\label{sn_sample}
\tablehead{

\multicolumn{2}{c}{designation} &
\multicolumn{2}{c}{host position} &
\multicolumn{1}{c}{stellar mass} &
\multicolumn{1}{c}{SFR}\\

\colhead{SN ID} &
\colhead{IAU} &
\colhead{$\alpha(J2000)$} &
\colhead{$\delta(J2000)$} &
\colhead{$[\log M_{\odot}]$} &
\colhead{$[\log M_{\odot}/yr]$} &
\colhead{SN$^a$} &
\colhead{Host$^b$} &
\colhead{Sample}
}
\startdata
1032  &  2005ez  & $03^h07^m11.016^s$ & $+01^{\circ}07^`11.96^{``}$ & $10.47^{+ 0.09}_{- 0.07}$ & $N/A$ &sp& p&  \\
1580  &  2005fb  & $03^h01^m17.544^s$ & $-00^{\circ}38^`38.63^{``}$ & $ 7.72^{+ 1.00}_{- 0.32}$ & $-0.98^{+ 1.02}_{- 0.36}$ &sp&sf& r  \\
15421  &  2006kw  & $02^h14^m57.912^s$ & $+00^{\circ}36^`09.80^{``}$ & $10.17^{+ 0.12}_{- 0.10}$ & $ 0.80^{+ 0.18}_{- 0.22}$ &sp&sf& r  \\
11172  & N/A & $21^h29^m39.120^s$ & $-00^{\circ}12^`07.88^{``}$ & $10.18^{+ 0.14}_{- 0.03}$ & $ 1.00^{+ 0.02}_{- 0.33}$ &lc&sf& r  \\
          &   & & & ... & & & &   \\
\enddata
\tablenotetext{a}{SN classification as a Ia based on spectra (sp) or light-curve shape (lc)}
\tablenotetext{b}{Our host classification either as passive (p) or star-forming (sf)}
\end{deluxetable}

\subsection{SDSS-II SN Sample}

In this analysis, we use the full dataset from the  SDSS-II Supernova Survey
\citep{2008AJ....135..338F}, which provides one of the largest samples of SNe Ia currently available. The SDSS-II SN Survey was a dedicated search for transient objects using the SDSS 2.5-m telescope and imaging camera 
\citep{2000AJ....120.1579Y, 2006AJ....131.2332G}
to perform repeat imaging of  the ``Stripe 82" region of the SDSS survey for 3 months a year from 2005 to 2007. The SDSS-II transient database contains many thousands of potential SN candidates, out to $z\sim0.5$, of which $\simeq$500 were spectroscopically confirmed as SNe Ia during the survey period \citep{2008AJ....135..348S, 2008AJ....136.2306H}. The first year (2005) of the SDSS-II SN sample was recently used for detailed cosmological analyses \citep{2009ApJS..185...32K, 2009ApJ...703.1374S, 2010MNRAS.401.2331L}.

For the host galaxy analysis presented herein, we focus on the low--redshift part ($z<0.21$) of the SDSS-II SN sample, where  the SN light curves are measured to high 
accuracy, with multiple, high signal-to-noise ratio data-points per light curve, and the k-corrections are empirically determined to be more reliable, i.e., minimizing the influence of the UV--part of the SN spectrum (see Foley et al. 2008; Kessler et al. 2009a). Furthermore, the efficiency of the SDSS-II SN survey remains above 50\% below this redshift limit as demonstrated in 
\citet{2010arXiv1001.4995D}.

To ensure our SN sample is more complete, we also include photometric SNe Ia that have a light curve consistent with being a Type Ia, based on the Bayesian light curve fitting of  \citet{2008AJ....135..348S}, and a known host galaxy spectroscopic redshift. The likely non-Ia contamination within these additional photometrically--classified Ia's is only $\simeq 3\%$ 
\citep{2010arXiv1001.4995D}.

In total, this provides a sample of 361 supernovae (for $z<0.21$), of which 258 are spectroscopically confirmed. We provide a breakdown of these SN numbers in Table \ref{tbl_sample} where ``Spec Confirm" gives just the spectroscopically confirmed SNe. 
In the following section, we describe the light curve fitting procedure and the host galaxy classification which lead to a further reduction in the available SN sample (see Table \ref{tbl_sample}).

\subsection{Fitting SN Light Curves}
\label{fitters}

Several algorithms are available for fitting the light curves of SNe, and determining cosmological distances and SN properties. The two most common, publicly--available, fitting methods are SALT2  \citep{2007A&A...466...11G} and MLCS2k2 \citep{2007ApJ...659..122J, 2009ApJS..185...32K}; we use both of these techniques to explore the dependencies of our results on the details of the light curve analysis.  For the main results of this paper, we use the public SALT2 light curve fitter and, in Appendix A, we provide a similar analysis using MLCS2k2. We find that the results and conclusions of this paper are consistent for both light curve fitting algorithms and different SN selection criteria.

For our main SALT2 light curve fitting analysis, we impose the following additional criteria to our SN sample (described in Section 2.1) to ensure robust measurements for the stretch and color of each SN (based on our experience with the first year SDSS-II SN cosmology analysis). First,  we require at least five epochs in all the SDSS $gri$ passbands, with at least one measurement before the light curve maximum. We also require that the reduced $\chi^2$ of the light curve fit to the data in each filter is less than three.  These cuts reject 108 SNe from our sample and are shown in Table \ref{tbl_sample}  labelled as  ``After LC cuts".

SALT2 reports for each individual
SN an apparent brightness ($m_B$) in the B-band, a stretch value ($x_1$) and a color (or $c$) term, which can then be used to calculate a distance modulus ($\mu$) using, 

\begin{equation}
\mu = (m_B - M)+\alpha x_1 - \beta  c,
\label{eq1}
\end{equation}

\noindent where $M$ is the ``standardized" absolute SN Ia magnitude (at $x_1=c=0$), $\alpha$ describes the overall stretch law for the sample and $\beta$ is the color law for the whole sample\footnote{If the color term is interpreted as solely due to dust absorption, then
the relationship $\beta = R_B = R_V + 1$ should hold. However, such an interpretation
is probably too simplistic due to intrinsic variations of the SN color.}. We only report the {\it uncalibrated} values of $M$ from SALT2 which are degenerate with our assumed value of $H_0$:  We are only interested in relative differences in the absolute brightness of SNe, not the true brightness. In most SN cosmological analyses, it is assumed that these parameters are invariant to the type of host galaxy in the 
sample and do not evolve with redshift, but there is no a-priori reason for such an assumption.

Based on the observed values of our SN stretch and color distributions, we remove a further 19 light curves by imposing the limits of $-4.5<x_1<2$ and $-0.3<c<0.6$, which are labelled in Table \ref{tbl_sample} as  ``After LC fitter limits". These limits were empirically determined to remove SN events where SALT2 reports extreme values either for $x_1$ or $c$
not resembling the majority of SN in our sample (see Fig. 2), leaving 234 SNe Ia for our main SALT2 analysis (as shown in Table 1). 

Following the usual SALT2 prescription, we determine $M$, $\alpha$ and $\beta$ in Eqn. \ref{eq1} by minimizing the scatter about a fiducal redshift-distance relation. We adopt as the reference cosmological model a flat universe with an energy density of matter of $\Omega_M=0.272$,
taken from \citet{2009ApJS..180..330K}, and $H_0=65\,{\rm km\,s^{-1}\,Mpc^{-1}}$. We have confirmed that the main results of this paper do not depend on the details of this assumed cosmological model. 

\begin{figure}[t]
\epsscale{.85}
\plotone{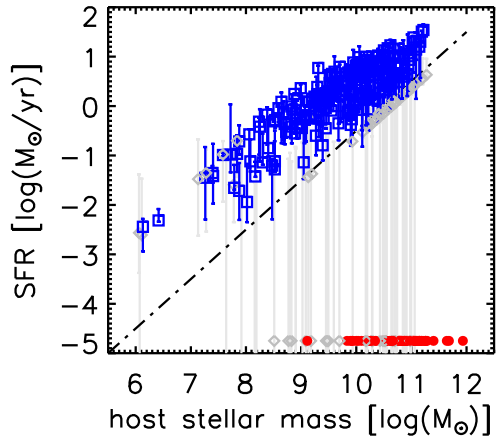}
\caption{
The distribution of stellar mass and star-formation rate (SFR) for the complete sample of host galaxies used herein calculated using P\'EGASE2. The (blue) open squares are star-forming host galaxies while solid (red) circles are passive hosts. The open (grey) diamonds are host galaxies excluded because either because of  their large error bar (shown in grey), which makes their classification less clear, or the fit to the SN light curve does not pass the selection cuts discussed in the text. The dashed line shows the  $\log(\textrm{sSFR}) < -10.5$ limit for star--forming galaxies discussed in the text. Passive galaxies (with zero SFR) are artificially plotted at $-4.75$ to be displayed in this figure.
\label{fig:pegase}}
\end{figure}

\subsection{Host Galaxy Properties}
\label{host_galaxy_properties}

A key part of our analysis is the host galaxy properties for each SN, which were determined using the techniques outlined in detail in Smith (2009). We begin by matching SN positions, within a 0.25 arcminute search radius, with SDSS galaxies detected in the deep optical stacked images of ``Stripe 82" constructed from the SDSS-I/II photometry \citep{2009ApJS..182..543A}
and choose the closest match as the host galaxy.  We also require that all SN host galaxies have a measured SDSS model magnitude of $r\le23$. These two constraints remove 7\% of our SNe with either a missing or too faint host. We then visually confirm each host, via inspection of images with and without the SN present, to ensure that the correct host has been associated with each SN.
In six cases, where the host is extended, or de-blended into multiple objects by the SDSS automated software, we
adjust the host position to the center of the underlying galaxy.

Each of the SNe Ia in our sample has a spectroscopic redshift, either from the SN itself or its host galaxy. This redshift is combined with the five SDSS photometric measurements (model magnitudes in $ugriz$ passbands corrected for Galactic extinction) for the host galaxy to determine the star-formation rate (SFR) and stellar mass of each system, using the P\'EGASE2 spectral energy distributions \citep{1997A&A...326..950F, 1999astro.ph.12179F} and the Z-PEG software package \citep{2002A&A...386..446L}. In detail, we used the eight star-forming scenarios, as listed in Table 1 of Le Borgne \& Rocca-Volmerange (2002), and assume a Kroupa (2001) initial mass function. In these scenarios, the SFR is defined for most galaxies via $\textrm{SFR} = \nu \times M_{\textrm{gas}}$, where $\nu$ (in units of $\textrm{Gyr}^{-1}$) ranges from 0.07 to 3.33, while for irregular galaxies, the SFR is defined as $\textrm{SFR} = 0.065 {M}_{\textrm{gas}}^{1.5}$ ($M_\textrm{gas}$ is the density of gas in solar masses). We use the default modeling of internal dust as discussed in Le Borgne \& Rocca-Volmerange (2002)
where a King profile is used for the elliptical template, whilst a plane-parallel slab distribution is used for spiral and irregular galaxies.
Each scenario is then evaluated at 69 different time-steps in their evolution, thus resulting in 552 possible galaxy template spectra covering a wide range of possible evolutionary scenarios. 

\begin{figure*}[t]
\epsscale{1.}
\plotone{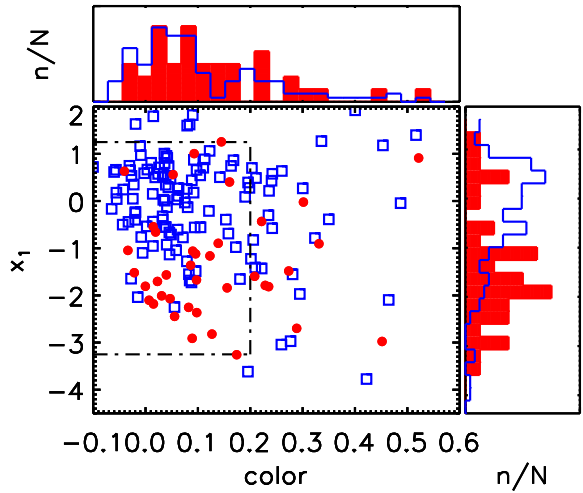}
\caption{
The observed distribution of the SALT2 $x_1$ and $c$ (color) values.
Red solid circles denote SNe in passive galaxies, whereas open blue squares indicate SNe in star-forming galaxies. The dot--dashed box shows the restricted subsample discussed in the text in Section \ref{testlaw}. The histograms in the top panel of the figure show the normalized distribution
in $c$ for the star-forming (blue open) and passive (red solid) host galaxies. The right--hand panel shows similar histograms but now for the $x_1$ distributions. \label{fig2}}
\end{figure*}

These templates were fit to the galaxy fluxes (converted from their model magnitudes after correcting them to the AB-system), with the redshift fixed to the redshift of the SN or host galaxy, to determine the best template for each host galaxy. The normalization is related to the total stellar mass of these templates, and is a free parameter which is determined as part of the fitting procedure to the host. We also use the best-fit template to estimate the recent star-formation rate of the host galaxy by integrating the best-fit template over the last half a gigayear of its evolution, e.g., if the best-fit template had an age of 8 Gyrs, then the recent SFR of the host was calculated over the period 8 to 7.5 Gyrs. Error bars on these estimates were determined by propagating the observed galaxy photometric errors. Our technique is similar to that used by Sullivan et al. (2006), and the Z-PEG software and the spectral energy distributions, have been used substantially in the literature (Glazebrook et al.2004; Grazian et al. 2006; Smith et al. 2010). 

To test the validity of our P\'EGASE2--based methodology, we have compared the stellar masses determined using the photometric data on over 350,000 SDSS Main galaxies (taken from DR4; Adelman-McCarthy et al. 2006) to masses derived by Kauffmann et al. (2004) using a technique based on SDSS spectral features. We find no mean difference between the two mass estimates, with a variance of only 3\%. 

In Figure \ref{fig:pegase}, we show the separation of host galaxies according to their stellar mass and SFR obtained from the P\'EGASE2 analysis above. We classify each host as either  passive (i.e., shows no sign of recent star-formation activity), or star-forming (i.e., having evidence for recent star-formation). To ensure a clean separation between these two galaxy classes, we require that the measured $1\sigma$ error on the estimated SFR for each galaxy is smaller than the separation in SFR for the two classes of galaxies, i.e., it is then unlikely that statistical errors on an individual SFR measurement can scatter a galaxy from one host galaxy type to the other. We also exclude star-forming host galaxies which have a specific SFR (i.e., sSFR; defined as the SFR per stellar mass) in the range $\log(\textrm{sSFR}) < -10.5$ as illustrated in Figure 1 as a dashed line. This limit excludes the locus of star--forming galaxies at the bottom of the blue cloud of points in Figure \ref{fig:pegase}, which are predominantly fit by the P\'EGASE2  lenticular galaxy scenario, thus leading to an unclear interpretation of their star--formation activity. These cuts and limits ensure we have two, well-separated, samples of host galaxies. We show in Table 1 the numbers of SNe available in these two host galaxy classes and note that many SNe have been excluded from further analysis because of the ambiguity of their host galaxy type. 
In Table \ref{sn_sample}, we provide the SN designation, host galaxy coordinates, host galaxy stellar mass and star-formation rate (derived from our best fit P\'EGASE2 model). We also provide whether the SN was spectroscopically confirmed (sp) or classified by just its light curve (lc), and if we classify the host galaxy as passive (p) or star-forming (sf).
More sophisticated stellar population models could have been used to determine the star--formation histories of our host galaxies \citep{2009A&A...493..425M} but we find that our classification of galaxies into two broad classes of star-formation activity is relatively unaffected by the choice of templates (see Smith 2009 for more details). Also, our host galaxy analysis is in good agreement with a simple cut on the color of the galaxies, e.g., dividing the galaxies at $u-r=2.22$ \citep{2001AJ....122.1861S}.

\begin{figure*}[t]
\epsscale{1.5}
\plotone{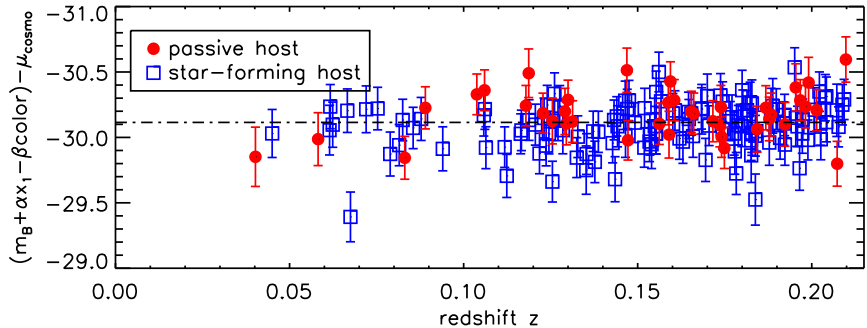}
\caption{The Hubble residuals as a function of redshift. The open blue squares denote SNe Ia in star-forming galaxies, while the solid red circles are SNe Ia in passive galaxies. The dashed line is the reference absolute magnitude fit to the whole sample regardless of host galaxy type. We find SNe Ia in passive host galaxies are $\simeq0.1$ magnitudes brighter than in star--forming hosts even after light curve fitting.
 \label{fig:hubbleresiduals}}
\end{figure*}

\section{Results}

\subsection{Testing the SALT2 Stretch and Color Relations}
\label{testlaw}

In Figure \ref{fig2}, we show the SALT2 output ($x_1$ and $c$ values from Eqn. 1) for our low redshift sample of SNe. The open (blue) squares represent SNe in host galaxies classified as star--forming as described in Section \ref{host_galaxy_properties}, whereas the solid (red) circles represent SNe in passive galaxies. In agreement with \citet{2006ApJ...648..868S}, we find a clear difference in the $x_1$ distributions between SNe Ia with a rapid decline rate (small $x_1$ values), which favor passive galaxies, and brighter, slower SNe (larger $x_1$ values) that favor star-forming galaxies. This result is clearly seen in the right-hand panel of Figure 2 and can be quantified using a Kolmogorov-Smirnov test (KS-test), where the probability for the two $x_1$ distributions being drawn from the same underlying parent distribution is only $10^{-7}$. This result has been known for some time \citep{{2000AJ....120.1479H}}, but it is reassuring that we can clearly recover this well-known difference in the $x_1$ distributions. 

However, we note that in Figure \ref{fig2} there is no clear separation in the color term ($c$) of SNe with respect to host galaxy type, i.e., both populations span the same range in color. This agreement is demonstrated in the top--panel of Figure 2 and a K-S test of the two $c$ distributions has a probability of $0.24$ of being drawn from the same underlying parent distribution, i.e., no evidence that they are drawn from different underlying distributions. This observation seems counter-intuitive, as we might expect some differences in the  global dust properties of these two host galaxy types, and maybe even an inclination dependence for the disk (star-forming) galaxies as outlined recently by \citet{2010arXiv1001.1744M}. This similarity in the color distributions implies that the rest-frame colors of SNe are dominated either by local, circumstellar dust, with the same color distributions, and/or SNe Ia have the same intrinsic color variations in all galaxy types.

\begin{deluxetable}{lcccccc}
\tabletypesize{\scriptsize}
\tablecaption{Best fit values for $M,\alpha, \beta$ as a function of host galaxy type}
\tablewidth{0pt}
\tablehead{
\colhead{Host Galaxies} & 
\colhead{Restricted$^a$} & 
\colhead{$M$} &
\colhead{$\alpha$} & 
\colhead{$\beta$} &
\colhead{$\chi^2$} &
\colhead{No. of SNe} 
}
\startdata
passive             &  no  & $-30.19\pm 0.03$ & $   0.16\pm 0.02 $ & $  2.42\pm 0.16 $ &    34.46  & 40   \\
                           &   yes  & $-30.23\pm 0.05$ & $   0.18\pm 0.03 $ & $  2.50\pm 0.41 $ &    12.60  & 27  \\
\hline
star-forming      & no &   $-30.10 \pm0.01$  & $ 0.12 \pm0.01$ &  $3.09 \pm0.10$  & 143.63   & 122  \\
                           & yes &  $-30.11 \pm0.02$  & $ 0.16 \pm0.02$ &  $3.22 \pm0.20$  &   94.55   &   89  \\
\hline
\enddata
\tablenotetext{a}{Restricted range in allowed $c$ and $x_1$ as shown in Figure 2.}
\end{deluxetable}

As discussed above, there is a clear trend for the $x_1$ distribution with 
host type, but no obvious trend for the color distribution. We explored if the constants in Eqn. 1 ($M$, $\alpha $, $\beta$) are dependent on host type by using a Markov chain Monte Carlo (MCMC) simulation
where we minimize the $\chi^2$ for the fit to the distance modulus versus redshift, as a function of $(M, \alpha , \beta)$ separately for passive and star--forming galaxies. Fits were obtained by running the MCMC chains with 50,000 accepted steps and adjusting the step size empirically to achieve a typical frequency for accepted steps of $\approx 20\%$.  One sigma errors on each parameter are provided by marginalizing over the remaining other parameters from the MCMC chains. We perform this analysis assuming an intrinsic dispersion of $\sigma_{int}=0.14$ mags, which is added in quadrature to the errors on the distance modulus to achieve a reduced $\chi^2$ close to one (i.e., $\chi^2/{\rm ndf} \approx 1$, see Lampeitl et al. 2010 for further discussion of this intrinsic dispersion).

To ensure our results are not driven by a few outliers, we also perform our analyses on a restricted subset of SNe with tighter allowed ranges of $c$ and $x_1$ values. This restricted sample is illustrated in Figure 2 as the inner dot-dashed box and reduces the sample from 162 SNe (see Table 1) to 116 SNe. 

In Table 2, we summarize our results for fitting $M,\alpha , \beta$ for the full sample and the restricted subsample discussed above. We see a correlation between the host galaxy type and the absolute magnitude ($M$) of the supernovae, i.e., after the SNe have been standardized using the SALT2 light curve fitting algorithm, there is still a difference of $\simeq0.1$ magnitudes, with SNe Ia in passive galaxies being brighter (more negative absolute magnitudes). To illustrate this effect, we present in Figure \ref{fig:hubbleresiduals} the residuals to the Hubble diagram (after removing the fiducial redshift--distance relation) for the best-fit SALT2 parameters of $(M,\alpha,\beta)=(-30.11, 0.12,2.86)$, which were determined from fitting the whole SN sample regardless of host galaxy type. As can be seen, there is a visible offset between SNe in passive and star--forming galaxies. 

The interpretation for the other SALT2 parameters ($\alpha $ and $\beta$) is less clear. First, we see no clear evidence for differences in $\alpha $ with host galaxy type given the statistical errors. Next, the fitted values of $\beta$ (the color law) for star-forming galaxies do appear to be larger than that found for passive galaxies, i.e., $\beta$ values for star-forming galaxies are above three, while for passive galaxies we find values below three. The significance of this difference in $\beta$ varies between the full and restricted samples,  which is not too surprising, as excluding the outliers in the color range will clearly increase the statistical error on the slope of the color law seen in Table 2. The mean slope ($\beta$) is similar for both the full and restricted sample. 

\begin{figure*}[t]
\epsscale{.99}
\plotone{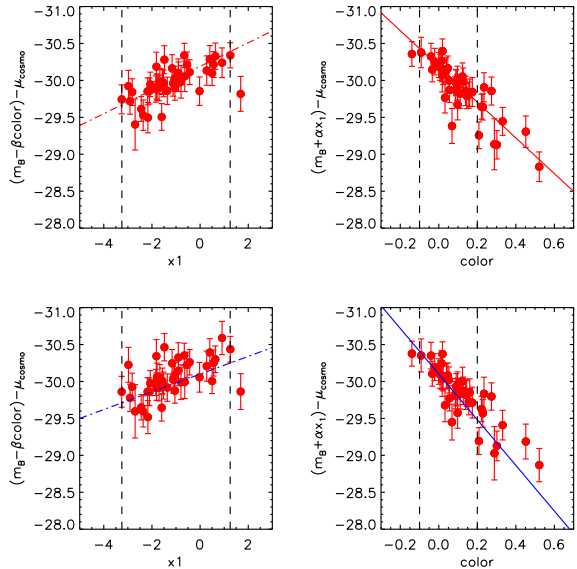}
\caption{(Left panels) The color--corrected absolute magnitudes of our SNe versus their $x_1$ parameters for our SN sample in passive host galaxies (without the $c$ and $x_1$ restrictions discussed in the text). (Right panels) The $x_1$-corrected absolute magnitudes versus color ($c$) of the SNe. In the upper row, we have applied to the data the best fit SALT2 parameters derived for the passive SN sample, whereas in the lower row we use the best fit SALT2 parameter for the star-forming SN sample. The vertical dashed lines indicate the restricted region in $x_1$ and $color$ discussed in Section 3.1 and shown as the inner dashed box in Figure \ref{fig2}. The inclined dashed line in the left plots indicates the applied 
stretch-correction ($\alpha  x_1$) and similarly the solid line in the right plots are the color--correction law ($\beta c$).
\label{fig3}}
\end{figure*}

In Figure \ref{fig3}, we show the corrected absolute magnitude for SNe in passive galaxies as a function of their fitted color and stretch values. The left-hand panels show the {\it color--corrected} absolute magnitude as a function of $x_1$, i.e., only the color part of Eqn. 1 ($\beta c$) has been applied to 
$m_B$. The right--hand panels show the {\it stretch--corrected} absolute magnitude, as a function of $c$, when only the stretch component of Eqn. 1 ($\alpha x_1$) has been applied to the distance modulus. In the upper panels, we show the best fitting law (Eqn. 1) assuming the best fit values of $M, \alpha, \beta$ for passive galaxies in Table 2, i.e., $(M, \alpha , \beta)_{P}=($-30.19$, 0.16, 2.42)$ respectively. In the lower row of panels, we show the best fit law again but now assuming the best fit parameters for star-forming galaxies, namely $(M, \alpha , \beta)_{SF} =($-30.10$, 0.12, 3.09)$. 

Comparing the top and bottom left--hand panels in Figure 4, it is clear we see that the amplitude ($M$) of the best-fitted relationship is different between the two and clearly wrong in the bottom panels (i.e., using the star-forming best-fit SALT2 parameters for SNe in passive galaxies). In Figure 5, we show the same analysis as in Figure 4, but this time the data plotted is for the star--forming SN sample. Again, we see that the amplitude of the fitted law (Eqn. 1) is different and inappropriate if used to describe the wrong type of galaxy.

\section{Discussion}

\subsection{Systematic Uncertainties}

Before we interpret these results, it is important to understand potential systematic uncertainties in our analysis. First, we have tested if our result depends on
the inclusion of a subset of photometrically confirmed SNe. We see negligible changes in the central values of $M_0$, $\alpha$ and $\beta$ (for SALT2) which are significantly 
smaller than the errors on these parameters. We also verify the robustness of our results to reasonable changes in the fiducial cosmological model and find no significant effect as expected. Likewise, we have increased the redshift range of the sample used in our analysis, e.g., increasing the limit to $z<0.45$, which more than doubles the size of the sample but makes the sample more incomplete. We find that the observed differences with host galaxy type are still present,  but the significance is decreased. This decrease in significance is likely caused by the decrease in signal-to-noise ratio for both the SN light curves and the galaxy photometry, as well as increases in the sample incompleteness (both spectral confirmations and Malmquist bias effects). The uncertainties in the k--corrections are also increased as the UV--part of the SN spectrum becomes more important. We note that Sullivan et al. (2010) sees similar results but for the higher redshift SNLS sample, thus suggesting that any decrease in significance we witness is probably caused by observational issues rather than evolution in the SN population.  For these reasons, we have chosen to focus on the cleanest, most efficient, sample of SDSS-II SN at $z<0.21$.

In Appendix A, we provide a parallel analysis of our data using the MLCS2k2 light curve fitting technique and find similar results to those seen in the SALT2 analysis in Section 3; namely differences in the absolute magnitude of SNe Ia between passive and star--forming host galaxies, as well as differences in the best-fit color laws  (as represented by $R_V$). This confirms that our results are not sensitive to the details of how the light curves were analysed and suggests the trends we see are either inherent to the supernovae, especially as we still see a correlation for the restricted sample of SN. However, further analysis is required to conclusively determine the fundamental origin of the observed correlations including potential improvements in the light curve fitting methodologies, e.g., a better representation of local SNe in passive hosts within the MLCS2k2 training sample or a more sophisticated parameterization of the $x_1$ and $c$ distributions in SALT2 to better accommodate the fast declining SNe Ia in passive hosts.

\begin{figure*}[t]
\epsscale{.99}
\plotone{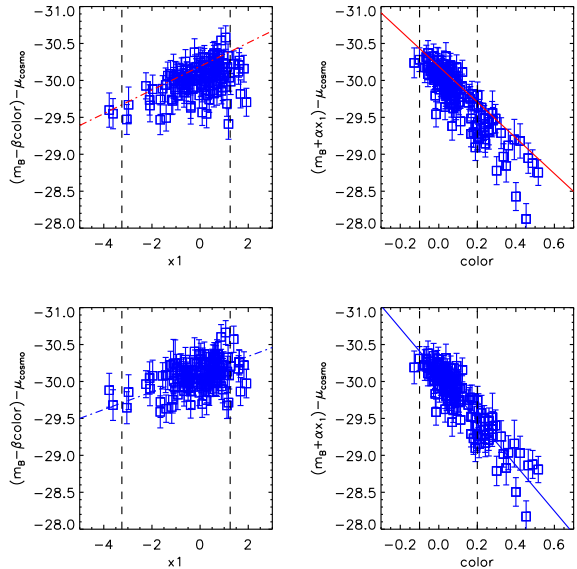}
\caption{Similar in Figure 4, but for SNe in star-forming host galaxies. The dashed lines are for the restricted subset of SNe discussed in Section 3.1.
\label{fig4}}
\end{figure*}

\subsection{Intrinsic Dispersion}

Throughout this analysis, we have assumed an intrinsic dispersion for our SN sample of $\sigma_{int}=0.14$ mags, which is consistent with the value obtained for the first year SDSS-II data analysed in Lampeitl et al. (2010). In Table 2 (and Table \ref{tab:mlcs}), we present the separate $\chi^2$ values for SNe in both passive and star-forming galaxies (for both SALT2 and MLCS2k2). The star--forming subsample has a reduced $\chi^2$ above unity, but for passive hosts, the reduced $\chi^2$ value is now less than one. This observation suggests 
that SNe in passive galaxies would favor a smaller intrinsic dispersion and are thus a more homogeneous population, or the observed errors are more representative of the scatter 
about the Hubble diagram.

To investigate this matter further, we have re--fit both the passive and star--forming SN samples (with no restrictions in $c$ and $x_1$) and adjusting the intrinsic dispersion 
$\sigma_{int}$
to a value that gives a reduced $\chi^2$ close to 1. In the case where we fit the passive sample with the parameters derived from just the passive sample we 
find $\sigma_{int}=0.13$, but 
these values give $\chi^2$ of 321 (for 122 SNe) for the star-forming sample. In the reverse, we fit the star-forming sample with the best-fit star-forming parameters and find 
$\sigma_{int}=0.17$, where now the passive sample yields $\chi^2=55$ for 40 SNe.
This result suggests that by using the larger $\sigma_{int}$ for SNe in passive hosts, we are effectively down-weighting these SNe (by increasing their errors) because of the offset in $M$ between the SNe Ia in the two types of galaxies.

\begin{figure*}[t]
\epsscale{1.}
\plotone{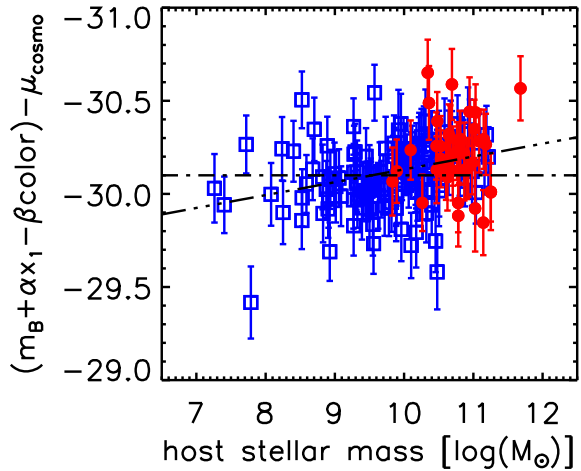}
\caption{The residuals around the best-fit Hubble diagram as a function of host galaxy stellar mass (calculated from P\'EGASE2). The (red) solid circles are for passive host galaxies, while the (blue) open squares are star-forming hosts. The inclined dot--dot--dashed line is the best--fit to these data, while the dot--dashed line is the fiducial 3--parameter SALT2 fitted model (Eqn. 1) without regard to the host galaxy stellar mass or type. 
\label{fig9}}
\end{figure*}

\subsection{Mass Dependency}

\begin{deluxetable}{lcccccccc}
\tabletypesize{\scriptsize}
\tablecaption{Best fit values $(M, \alpha, \beta, \gamma)$ \label{tab:gamma}}
\tablewidth{0pt}
\tablehead{
\colhead{Dataset} & 
\colhead{M} &
\colhead{$\alpha$} & 
\colhead{$\beta$} &
\colhead{$\gamma$} &
\colhead{$\chi^2$} &
\colhead{No. SNe} 
}
\startdata
Full SN sample &$ -30.09\pm0.01$ & $0.13\pm0.01$ & $3.04\pm0.07$ & $0.072\pm0.018$ & $ 179.53$ & $ 162$  \\
Restricted & $ -30.11\pm0.01$ & $0.16\pm0.02$ & $3.22\pm0.30$ & $0.088\pm0.008$ & $  98.87$ & $ 116$  \\
 
\hline
\enddata
\end{deluxetable}

As discussed in Section 2.3, we have classified our SN host galaxies into two well--separated classes, namely passive and star-forming. We initially separated the galaxies in this way because previous studies of the properties, and rates, of SNe Ia have shown clear correlations with the star--formation activity of the host, e.g., Hamuy et al. (2000), Sullivan et al. (2006), Manucci et al. (2005), Dilday et al. (2010b). Gallagher et al. (2008) showed that the measured metallicity for local SN host galaxies was correlated to the residuals around the best fit distance-redshift relation, which has prompted some authors to look for correlation with the host galaxy stellar mass as a proxy for the metallicity (using the known mass-metallicity relationship; Tremonti et al. 2004). Both Kelly et al. (2009) and Sullivan et al. (2010) find such a correlation with stellar mass and Sullivan et al. (2010) exploit this correlation to improve the cosmological fits to the 3-year SNLS dataset.

We present here a first analysis of the SDSS-II Hubble residuals as a function of the host galaxy stellar mass. The advantage of such an analysis is that the host stellar mass is a continuous parameter thus potentially avoiding uncertainties associated with binning galaxies into two distinct samples. Unfortunately, estimating the stellar mass from broad-band photometry is challenging and therefore, the measurements of stellar mass can be noisy and potentially biased. For the analysis below, we therefore restrict ourselves to the clean sample defined in Section 2.3, i.e., we still do not include galaxies with an ambiguous classification between star-forming and passive.

 In Figure \ref{fig9}, we show the Hubble residuals as a function of stellar mass. As expected, the passive galaxies have preferentially higher stellar masses than the star-forming subsample. A linear fit to the Hubble residuals is shown as the inclined line (dot-dot-dashed) and has a slope of $0.069\pm0.014$ magnitudes per $\rm{log(M_{\sun})}$. We see the same result, at the same statistical significance, with our MLCS2k2 analysis in Appendix A. Therefore, these results imply that adding an additional parameter, dependent on the host stellar mass, to Eqn. \ref{eq1} would give a better fit to the SDSS-II SN Hubble diagram. 
 
To quantify this statement, we have repeated our MCMC analysis in Section 3.1, but now minimizing over 4 parameters $(M, \alpha, \beta, \gamma)$ where calculating the distance modulus using,
\begin{equation}
\mu = (m_B - M)+\alpha x_1 - \beta  c+\gamma m_{st},
\label{eq2}
\end{equation}
where $m_{st}$ is defined as
\begin{equation}
m_{st} = \log(m_{host}) - 9.5,
\label{log:hostmass}
\end{equation}
and $m_{host}$ is the stellar mass of the host in units of solar mass derived from our P\'EGASE2 fits to the host galaxy colors (Section 2.3).

The results of this new analysis are presented in Table \ref{tab:gamma}. First, we see that adding an additional parameter has not significantly changed the fitted values of $M$, $\alpha$ and $\beta$; they are close to the values in Table 2 for star-forming hosts (which dominate the whole sample). Secondly, we see a significant non-zero value for $\gamma$ at $>4\sigma$. Finally, we can compare the three and four-parameter fits (Eqns. 1 and 2 respectively) to the full SN sample, regardless of host galaxy type, using the Bayesian Information Criteria (BIC; Liddle 2004). BIC is a penalized likelihood statistic that accounts for models with different numbers of parameters, and we find the BIC score for Eqn. 2 is 117 compared to 134 for Eqn. 1. The smaller BIC score demonstrates the additional parameter is justified. 

\section{Conclusions}

We present an analysis of the host galaxy dependencies for the SDSS-II Supernova Survey.  We have used 361 SNe Ia (see Table \ref{tbl_sample}) taken from the full three years of this survey, and then applied several data cuts to ensure we have a clean, well-understood, sample of low redshift SNe ($z<0.21$). We have analysed these data using two well-known light--curve fitting routines (SALT2 and MLCS2k2) to demonstrate that our results are not dependent on the details of the light curve analysis. We summarise below the main conclusions of this work:

\begin{itemize}

\item We confirm, to high significance, the strong correlation between host galaxy type and the observed width of the light curve, i.e., quick decline--rate SNe (small $x_1$ values in SALT2), favor passive host galaxies, while bright, slower decline SNe Ia (larger $x_1$ values) favor star-forming galaxies. This has been seen before by several authors. However, we find no correlation between the color of individual SNe Ia and their host galaxy, as illustrated in Figure 2. 

\item We find that SNe Ia are $\simeq0.1$ magnitudes brighter in passive host galaxies after light curve fitting. This effect is true for both SALT2 and MCLS2k2 analyses. The statistical significance of this difference is between $2$ and $3\sigma$ dependent upon the details of the fitting methodology and the inclusion of outliers in the color and $x_1$ distributions of these data. 

\item We find evidence for differences in the SN color relationship between passive and star--forming host galaxies. For SALT2, we detect differences in $\beta$, with passive hosts showing $\beta\simeq2.5$ and star--forming hosts prefering $\beta>3$. For MLCS2k2, we see a similar trend for passive hosts preferring a dust law with $R_V\simeq1$ and star-forming hosts giving $R_V\sim2$. The significance of these trends depends on the color range considered, but is greater than $3\sigma$ for the full SN sample considered herein.

\item We find that the required intrinsic dispersion for passive galaxy hosts is smaller than that needed for the whole SN sample (and for star--forming hosts), e.g., only 
$\sigma_{int}=0.13$ mags is required to obtain a reduced $\chi^2$ close to unity for passive hosts compared to 0.17 mags for the star-forming sample. This lower intrinsic 
dispersion for passive hosts is true for both SALT2 and MLCS2k2 light curve fitters.

\item We demonstrate that the dependence on host galaxy type can be parameterized using the stellar mass of the host galaxies. We show that a 4--parameter fit to the distance modulus of SNe Ia ($M$, $\beta$, $\alpha$, $\gamma$) -- where $\gamma$ scales with stellar mass -- is better than the usual 3--parameter model given in Eqn. 1. For the data in Figure \ref{fig9}, we find $\gamma=0.069\pm0.014$, or a $4\sigma$ detection of this parameter. 
\end{itemize}

These conclusions are in good agreement with other work, especially Kelly et al. (2009) and Sullivan et al. (2010). In particular, Sullivan et al. (2010) see the same trends in both $M$ and $\beta$ discussed herein with a similar level of statistical significance. This indicates that these trends are common to several SN surveys and appear to not change significantly with redshift. One possible cause for these correlations is a difference in the host galaxy metallicity (see Gallagher et al. 2008), which is correlated with the host stellar mass and host star-formation activity, and could affect the metallicity of the progenitor star thus leading to changes in the peak brightness of SNe Ia. However, the origin of these correlations requires further study, especially to ensure deficiencies in the light curve fitting techniques are not directly responsible. 

The host galaxy dependencies presented in this paper will be important for future supernova cosmology surveys, which may wish to exploit these dependencies to minimize the scatter on the SN Hubble diagram. This could be achieved by including further parameters in the light curve fitting or distance modulus calculation (e.g., Eqn. 2). We will explore these issues in the future with the full SDSS-II SN Survey similar to the recent analysis of Sullivan et al. (2010) for the SNLS.

\acknowledgments

\section*{Acknowledgements}

The authors thank the referee for their careful reading of the paper. We also like to thank Mark Sullivan, Janine Pforr and Isobel Hook for helpful discussions during the course of this research. We thank Rick Kessler for his comments on an earlier draft of this paper. HL and RCH were supported by STFC, while MS is funded by an SKA fellowship.
MI acknowledges the support from the grant No. 2009-0063616 from MEST. SWJ acknowledges support from US Dept. of Energy grant DE-FG02-08ER41562.

Funding for the creation and distribution of the SDSS and SDSS-II
has been provided by the Alfred P. Sloan Foundation,
the Participating Institutions,
the National Science Foundation,
the U.S. Department of Energy,
the National Aeronautics and Space Administration,
the Japanese Monbukagakusho,
the Max Planck Society, and the Higher Education Funding Council for England.
The SDSS Web site \hbox{is {\tt http://www.sdss.org/}.}

The SDSS is managed by the Astrophysical Research Consortium
for the Participating Institutions.  The Participating Institutions are
the American Museum of Natural History,
Astrophysical Institute Potsdam,
University of Basel,
Cambridge University,
Case Western Reserve University,
University of Chicago,
Drexel University,
Fermilab,
the Institute for Advanced Study,
the Japan Participation Group,
Johns Hopkins University,
the Joint Institute for Nuclear Astrophysics,
the Kavli Institute for Particle Astrophysics and Cosmology,
the Korean Scientist Group,
the Chinese Academy of Sciences (LAMOST),
Los Alamos National Laboratory,
the Max-Planck-Institute for Astronomy (MPA),
the Max-Planck-Institute for Astrophysics (MPiA), 
New Mexico State University, 
Ohio State University,
University of Pittsburgh,
University of Portsmouth,
Princeton University,
the United States Naval Observatory,
and the University of Washington.

This work is based in part on observations made at the 
following telescopes.
The Hobby-Eberly Telescope (HET) is a joint project of the University of Texas
at Austin,
the Pennsylvania State University,  Stanford University,
Ludwig-Maximillians-Universit\"at M\"unchen, and Georg-August-Universit\"at
G\"ottingen.  The HET is named in honor of its principal benefactors,
William P. Hobby and Robert E. Eberly.  The Marcario Low-Resolution
Spectrograph is named for Mike Marcario of High Lonesome Optics, who
fabricated several optical elements 
for the instrument but died before its completion;
it is a joint project of the Hobby-Eberly Telescope partnership and the
Instituto de Astronom\'{\i}a de la Universidad Nacional Aut\'onoma de M\'exico.
The Apache 
Point Observatory 3.5 m telescope is owned and operated by 
the Astrophysical Research Consortium. We thank the observatory 
director, Suzanne Hawley, and site manager, Bruce Gillespie, for 
their support of this project.
The Subaru Telescope is operated by the National 
Astronomical Observatory of Japan. The William Herschel 
Telescope is operated by the 
Isaac Newton Group on the island of La Palma
in the Spanish Observatorio del Roque 
de los Muchachos of the Instituto de Astrofisica de 
Canarias. The W. M. Keck Observatory is operated as a scientific partnership 
among the California Institute of Technology, the University of 
California, and the National Aeronautics and Space Administration; the 
observatory was made possible by the generous financial support of the 
W. M. Keck Foundation.

\begin{deluxetable}{lcccccc}
\tabletypesize{\scriptsize}
\tablecaption{Best fit values for $R_V$ and $H_0$ as a function of host galaxy type using MLCS2k2 \label{tab:mlcs}}
\tablewidth{0pt}
\tablehead{
\colhead{Dataset} &
\colhead{$A_V$ range} &
\colhead{$R_V$} &
\colhead{$H_0$} &
\colhead{$\chi^2$} &
\colhead{No. SNe}
}
\startdata
Passive  & Full & ${1.0}^{+0.5}_{-0.1}$ & $66.67 \pm 0.94$ & 21.53 & 35  \\  %%here
	         & [0,1] & $1.0 \pm 0.2$ & $66.88 \pm 0.95$ & 19.82 & 33 \\ \hline
Star-forming& Full & $1.8 \pm 0.1$ & $62.75 \pm 0.50$ & 90.08 & 100  \\%%here
		 & [0,1] & ${1.8}^{+0.2}_{-0.4}$ & $62.97 \pm 0.53$ & 63.22 & 89  \\
\hline
\hline
Passive	& Full &  ${1.6}^{+0.1}_{-0.2}$ & $68.63 \pm 0.95$ & 25.03 & 35  \\
(No Prior)	 & [-0.5,1] & ${1.0}^{+0.4}_{-0.3}$ & $66.73 \pm 1.11$ & 16.80 & 24 \\
\hline
Star-forming & Full & $1.7 \pm 0.1$ & $62.47 \pm 0.49$ & 80.05 & 100  \\
(No Prior)	& [-0.5,1] & ${1.6}^{+0.3}_{-0.1}$ & $62.30 \pm 0.53$ & 60.69 & 85  \\
\hline
\enddata
\end{deluxetable}

\section{Appendix A: MLCS2k2 analysis}

\subsection{Parameter and Sample Selection}

In addition to the SALT2 light-curve fitter in Section 2.2, we also provide a parallel analysis based on the MLCS2k2 SNANA light curve technique (Jha et al. 2007; Kessler et al. 2009a,b). In this way, we can determine if the correlations seen in Section \ref{testlaw} are just an artifact of the light curve analysis. As in Section 2.2, we assume a flat cosmology of $\Omega_M=0.272$ from \citet{2009ApJS..180..330K}, as well as assume the true absolute magnitude of SNe Ia is $M=-19.44$ (Kessler et al. 2009a) and an intrinsic dispersion of $\sigma_{int}=0.14$ mags.

In SALT2, the parameters $\alpha$ and $\beta$ describe the global stretch and color laws of SNe Ia and are determined through minimising the scatter around a cosmological model. MLCS2k2 takes a different approach in that the distance is both linearly and quadratically dependent on the light curve decline rate parameter ($\Delta$) compared to the purely linear dependence of $x_1$ in SALT2. The global correction (analogous to $\alpha$ in SALT2) is determined from a well-measured, low-redshift training set, prior to fitting. Any excess color variation is assumed to be due to extinction by dust in the host galaxy, and is parameterized using the \citet{1989ApJ...345..245C} law, where the excess is $E(B-V) = {A}_{V} / {R_V}$. For our Galaxy, $R_V = 3.1$, whilst previous SN Ia studies favor values of $R_V \sim 2.1$ (see references in Section 1). 

In this analysis, we attempt to constrain $R_V$ as a function of host galaxy type. MLCS2k2 does not minimise the scatter on the Hubble diagram, but assumes a value for $R_V$ in the fitting process. Thus, we allow $R_V$ to vary between $0.1 < R_V < 4.0$, in increments of 0.1, and minimise the $\chi^2$ of the fit to the Hubble diagram to determine the best-fitting $R_V$ value. For $A_V$, we assume a ``standard" prior distribution in the fitting process as outlined in \citet{2009ApJS..185...32K}, who used $P({A}_{V}) = \exp{(-{A}_{V}/\tau)}$ with $\tau = 0.33$, based on the first year SDSS-II data. To quantify the effect of this assumption on our results, we also consider the case where no prior on the $A_V$ distribution is assumed, thus allowing it to take any value, and mimicking the SALT2 approach. 

As with SALT2, and following the methodology of Kessler et al. (2009a,b), we require that each light curve in our MLCS2k2 analysis has at least five photometric observations in the SDSS $gri$ passbands located between $-20$ and $+60$ days relative to maximum light in the SN rest frame. Of these epochs, at least one must be two or more days prior to maximum light, and at least one must be ten or more days past maximum light$\footnote{This additional cut compared to the SALT2 analysis in Section 3 only affects 17 SNe and has no impact on the conclusions of this paper}$, with at least one epoch with a $S/N > 5$ in each of the passbands (although not necessarily the same epoch each time). These criteria ensure we have well-measured light curves for 256 SNe at $z<0.21$ (see Table 1 for details).

In addition, we impose further constraints on the determined light curve shape and fit probability. We follow the procedure of \citet{2009ApJS..185...32K} and remove any object with a low probability of being a SN Ia (i.e. those with $P({I}_{a}) < {10}^{-3}$), or which has an unphysical value of $\Delta < -0.4$. These cuts remove a further 42 events (Table 1). As the value of $A_V$ is dependent on $R_V$, we require that each SN satisfies the criteria above for all values of $R_V$ used. Using the standard $A_V$ prior above, we are left with 214 SNe Ia (labelled as ``After LC fitter limits" in Table 1). To test the robustness of our results to outliers, we also restrict the range of $A_V$ values allowed (to $0.0 < A_V < 1.0$), which would reduce the sample to 198 SNe. Finally, we also apply the same cuts on the host galaxy classifications as described in Section 2.3 and presented as ``Valid host galaxy type" in Table 1.

\subsection{MLCS2k2 Results}

\begin{figure}[t]
\epsscale{.85}
\plotone{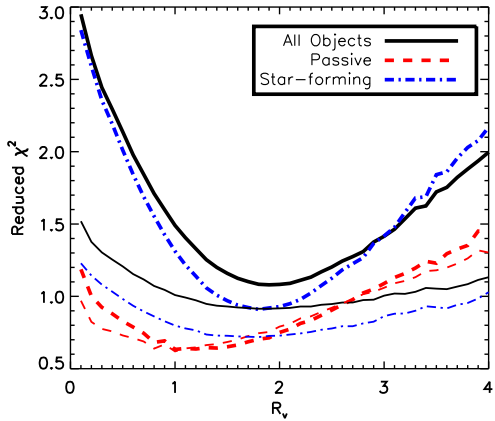}
\caption{The reduced $\chi^2$ as a function of $R_V$ from our MLCS2k2 analysis. The dashed (red) lines represent SNe in passive galaxies, while the dashed-dotted (blue) lines are for SNe in star--forming galaxies and the solid lines for the full sample. Thin lines indicate the results for the restricted sample with $0<A_V < 1.0$.
\label{fig7}}
\end{figure}

In Table~\ref{tab:mlcs}, we present the best fitting values of $R_V$ and $H_0$ for our MLCS2k2 analysis as well as their one sigma errors, which we calculated from the $\chi^2_{minimum}+1$ range holding the other parameters constant at their best-fit values. In the case where the error on $R_V$ can not be determined because it is smaller than the discrete binned values of the $R_V$ parameter, we then conservatively assume a error of $0.1$ (the spacing between the $R_V$ grid points). The top of Table~\ref{tab:mlcs} presents results assuming the standard $A_V$ prior distribution  and we see a clear preference for higher values of Hubble Constant for SNe in passive galaxies compared to star--forming galaxies (and the whole sample together). This preference is seen regardless of the $A_V$ range allowed. 

This observed difference in Hubble Constant can be due to differences in the assumed absolute magnitude ($\Delta M$) of SNe in different galaxy types, and we can convert between the two parameters using
\begin{equation}
\Delta M = 5 \log_{10} (\Delta{H}_{0}),
\label{eq4}
\end{equation}
where $\Delta H_0$ is the difference in Hubble Constants between the two samples. 
Therefore, in Table \ref{tab:mlcs}, we see $\Delta H_0=3.1 \textrm{km}\, \textrm{s}^{-1} \textrm{Mpc}^{-1}$ (assuming the standard $A_V$ prior distribution and the full SN sample), which translates to a $\simeq 0.12$ magnitudes difference (using Eqn. \ref{eq4}), with SNe in passive galaxies being brighter. 
This result is consistent with our SALT2 analysis in Section \ref{testlaw} and demonstrates that this result is independent of the details of the analysis. 

Table~\ref{tab:mlcs} also contains the best-fitting $R_V$ values, and we see a difference between the $R_V$ laws for SNe in the passive and star--forming galaxy samples: Supernovae in passive galaxies appear to favor $R_V\simeq1$ compared to $R_V\sim2$ in star--forming galaxies. This difference in $R_V$, as a function of host galaxy type, could be worrying for SN cosmology analyses which typically use a fixed value of $R_V$ for all SNe and often constrain the allowed range to  $1.7 < R_V < 2.5$. This range is consistent with our star--forming SN samples (and the whole sample), but inconsistent with our passive galaxy SN sample. 

In Figure~\ref{fig7}, we show how the reduced ${\chi^2}$ of our fits to the Hubble diagram varies as a function of $R_V$ for our sample. The solid lines indicate the $R_V$ values when all values of $A_V$ are allowed, while the dashed lines are only using SNe Ia with $0 <A_V <1$. This figure re--enforces the results in Table~\ref{tab:mlcs} in that there is a difference in the minina of these curves for the different galaxy types. As with the SALT2 analysis in Section \ref{testlaw}, we see that constraining the color range allowed decreases the difference as we are removing outliers to the main color laws.

The bottom row of Table~\ref{tab:mlcs} shows the results assuming no prior distribution on $A_V$ during the MLCS2k2 light-curve fitting, i.e., any $A_V$ value is allowed. In theory, this should be the closest match to the SALT2 analysis presented in Section \ref{testlaw}. Again, we see differences in the fitted $H_0$ values between the passive and star--forming host galaxy samples. Intriguingly, the evidence for a difference in $R_V$ between the two host galaxy types is now less, which is  consistent with the SALT2 results in Section \ref{testlaw} where we only witnessed a slight dependence on $\beta$ for the galaxy type. Clearly the choice of $A_V$ prior, as well as the allowed range of $A_V$ values, can have a significant effect on the best--fitting color parameters. 

\newpage

\begin{deluxetable}{llrrrrrrrr}

\tabletypesize{\scriptsize}
\setcounter{table}{1}
\tablecaption{For publication as supplement material}
\tablewidth{0pt}
\tablehead{

\multicolumn{2}{c}{designation} &
\multicolumn{2}{c}{host position} &
\multicolumn{1}{c}{stellar mass} &
\multicolumn{1}{c}{SFR}\\

\colhead{SN ID} &
\colhead{IAU} &
\colhead{$\alpha(J2000)$} &
\colhead{$\delta(J2000)$} &
\colhead{$[\log M_{\odot}]$} &
\colhead{$[\log M_{\odot}/yr]$} &
\colhead{SN} &
\colhead{Host} &
\colhead{Sample}
}
\startdata
1032  &  2005ez  & $03^h07^m11.016^s$ & $+01^{\circ}07^`11.96^{``}$ & $10.47^{+ 0.09}_{- 0.07}$ & $N/A$ &sp& p&   \\
1241  &  2005ff  & $22^h30^m41.040^s$ & $-00^{\circ}46^`34.47^{``}$ & $10.52^{+ 0.18}_{- 0.21}$ & N/A &sp& p& r \\
1371  &  2005fh  & $23^h17^m29.760^s$ & $+00^{\circ}25^`46.83^{``}$ & $10.76^{+ 0.21}_{- 0.10}$ & N/A &sp& p& r \\
2308  &  2005ey  & $02^h17^m05.616^s$ & $+00^{\circ}16^`50.88^{``}$ & $10.26^{+ 0.07}_{- 0.01}$ & N/A &sp& p&   \\
2689  &  2005fa  & $01^h39^m36.000^s$ & $-00^{\circ}45^`28.65^{``}$ & $11.15^{+ 0.22}_{- 0.10}$ & N/A &sp& p&   \\
12781  &  2006er  & $00^h21^m37.886^s$ & $-01^{\circ}00^`38.20^{``}$ & $10.78^{+ 0.24}_{- 0.09}$ & N/A &sp& p& r \\
14421  &  2006ia  & $02^h07^m19.176^s$ & $+01^{\circ}15^`07.24^{``}$ & $11.25^{+ 0.07}_{- 0.01}$ & N/A &sp& p& r \\
14782  &  2006jp  & $20^h56^m56.160^s$ & $-00^{\circ}16^`44.99^{``}$ & $11.03^{+ 0.19}_{- 0.17}$ & N/A &sp& p& r \\
14816  &  2006ja  & $22^h26^m51.840^s$ & $+00^{\circ}30^`23.08^{``}$ & $10.48^{+ 0.07}_{- 0.02}$ & N/A &sp& p& r \\
15201  &  2006ks  & $22^h30^m04.560^s$ & $+00^{\circ}00^`11.31^{``}$ & $11.14^{+ 0.08}_{- 0.10}$ & N/A &sp& p&   \\
15222  &  2006jz  & $00^h11^m24.578^s$ & $+00^{\circ}42^`07.30^{``}$ & $11.18^{+ 0.38}_{- 0.02}$ & N/A &sp& p& r \\
15648  &  2006ni  & $20^h54^m52.560^s$ & $-00^{\circ}11^`44.92^{``}$ & $11.08^{+ 0.07}_{- 0.10}$ & N/A &sp& p&   \\
15897  &  2006pb  & $00^h46^m43.584^s$ & $-01^{\circ}01^`56.96^{``}$ & $10.61^{+ 0.18}_{- 0.16}$ & N/A &sp& p& r \\
16206  &  2006pe  & $00^h23^m09.194^s$ & $-00^{\circ}03^`12.85^{``}$ & $10.67^{+ 0.17}_{- 0.01}$ & N/A &sp& p& r \\
16392  &  2006ob  & $01^h51^m48.504^s$ & $+00^{\circ}15^`49.81^{``}$ & $11.25^{+ 0.29}_{- 0.05}$ & N/A &sp& p& r \\
16641  &  2006pr  & $01^h34^m14.784^s$ & $-00^{\circ}24^`13.23^{``}$ & $10.47^{+ 0.05}_{- 0.31}$ & N/A &sp& p& r \\
17886  &  2007jh  & $03^h36^m01.584^s$ & $+01^{\circ}06^`17.14^{``}$ & $11.03^{+ 0.09}_{- 0.06}$ & N/A &sp& p&   \\
18298  &  2007li  & $01^h13^m04.032^s$ & $-00^{\circ}32^`24.01^{``}$ & $10.69^{+ 0.10}_{- 0.07}$ & N/A &sp& p&   \\
18415  &  2007la  & $22^h29^m54.720^s$ & $+01^{\circ}03^`30.49^{``}$ & $10.87^{+ 0.14}_{- 0.11}$ & N/A &sp& p& r \\
18604  &  2007lp  & $22^h43^m41.040^s$ & $+00^{\circ}25^`13.84^{``}$ & $10.78^{+ 0.11}_{- 0.02}$ & N/A &sp& p& r \\
18749  &  2007mb  & $00^h50^m11.184^s$ & $+00^{\circ}40^`32.56^{``}$ & $10.87^{+ 0.07}_{- 0.01}$ & N/A &sp& p& r \\
18809  &  2007mi  & $03^h23^m31.344^s$ & $+00^{\circ}40^`02.18^{``}$ & $10.90^{+ 0.09}_{- 0.01}$ & N/A &sp& p& r \\
18835  &  2007mj  & $03^h34^m44.496^s$ & $+00^{\circ}21^`19.85^{``}$ & $10.53^{+ 0.01}_{- 0.01}$ & N/A &sp& p& r \\
19174  &  2007or  & $01^h42^m38.352^s$ & $+01^{\circ}01^`49.22^{``}$ & $10.80^{+ 0.22}_{- 0.01}$ & N/A &sp& p& r \\
20048  &  2007pq  & $22^h37^m13.920^s$ & $+00^{\circ}44^`10.73^{``}$ & $10.63^{+ 0.14}_{- 0.07}$ & N/A &sp& p& r \\
20064  &  2007om  & $23^h54^m20.640^s$ & $-00^{\circ}55^`02.10^{``}$ & $11.02^{+ 0.19}_{- 0.30}$ & N/A &sp& p& r \\
20376  &  2007re  & $21^h17^m35.040^s$ & $-00^{\circ}31^`26.28^{``}$ & $10.34^{+ 0.07}_{- 0.01}$ & N/A &sp& p&   \\
1580  &  2005fb  & $03^h01^m17.544^s$ & $-00^{\circ}38^`38.63^{``}$ & $ 7.72^{+ 1.00}_{- 0.32}$ & $-0.98^{+ 1.02}_{- 0.36}$ &sp&sf& r \\
2031  &  2005fm  & $20^h48^m10.320^s$ & $-01^{\circ}10^`16.93^{``}$ & $ 9.28^{+ 0.26}_{- 0.34}$ & $-0.10^{+ 0.70}_{- 0.07}$ &sp&sf& r \\
2440  &  2005fu  & $02^h50^m32.136^s$ & $+00^{\circ}48^`26.44^{``}$ & $10.40^{+ 0.08}_{- 0.18}$ & $ 1.02^{+ 0.01}_{- 0.23}$ &sp&sf& r \\
2635  &  2005fw  & $03^h30^m48.960^s$ & $-01^{\circ}14^`15.40^{``}$ & $ 9.91^{+ 0.18}_{- 0.01}$ & $ 0.73^{+ 0.02}_{- 0.33}$ &sp&sf& r \\
2992  &  2005gp  & $03^h41^m59.352^s$ & $-00^{\circ}46^`58.51^{``}$ & $ 9.95^{+ 0.47}_{- 0.06}$ & $-0.01^{+ 0.74}_{- 0.03}$ &sp&sf&   \\
3087  &  2005gc  & $01^h21^m37.584^s$ & $-00^{\circ}58^`38.01^{``}$ & $ 9.45^{+ 0.01}_{- 0.04}$ & $ 0.25^{+ 0.15}_{- 0.01}$ &sp&sf& r \\
3256  &  2005hn  & $21^h57^m04.080^s$ & $-00^{\circ}13^`24.45^{``}$ & $ 9.72^{+ 0.31}_{- 0.19}$ & $ 0.23^{+ 0.21}_{- 0.56}$ &sp&sf& r \\
3317  &  2005gd  & $01^h47^m51.048^s$ & $+00^{\circ}38^`25.80^{``}$ & $ 9.86^{+ 0.01}_{- 0.06}$ & $-0.10^{+ 0.01}_{- 0.01}$ &sp&sf& r \\
3592  &  2005gb  & $01^h16^m12.600^s$ & $+00^{\circ}47^`30.90^{``}$ & $ 8.94^{+ 0.17}_{- 0.09}$ & $-0.43^{+ 0.29}_{- 0.22}$ &sp&sf& r \\
3901  &  2005ho  & $00^h59^m24.096^s$ & $+00^{\circ}00^`09.44^{``}$ & $ 9.73^{+ 0.09}_{- 0.15}$ & $ 0.24^{+ 0.35}_{- 0.03}$ &sp&sf& r \\
5103  &  2005gx  & $23^h59^m32.160^s$ & $+00^{\circ}44^`12.91^{``}$ & $ 9.29^{+ 0.00}_{- 0.00}$ & $ 0.24^{+ 0.00}_{- 0.01}$ &sp&sf& r \\
5395  &  2005hr  & $03^h18^m33.816^s$ & $+00^{\circ}07^`24.03^{``}$ & $ 8.88^{+ 0.01}_{- 0.01}$ & $ 0.24^{+ 0.01}_{- 0.01}$ &sp&sf& r \\
5550  &  2005hy  & $00^h14^m23.602^s$ & $+00^{\circ}19^`59.09^{``}$ & $ 9.31^{+ 0.22}_{- 0.08}$ & $ 0.77^{+ 0.05}_{- 0.68}$ &sp&sf&   \\
5635  &  2005hv  & $22^h12^m43.920^s$ & $-00^{\circ}02^`06.14^{``}$ & $ 9.56^{+ 0.01}_{- 0.11}$ & $ 0.65^{+ 0.04}_{- 0.55}$ &sp&sf& r \\
5944  &  2005hc  & $01^h56^m47.976^s$ & $-00^{\circ}12^`48.62^{``}$ & $ 7.26^{+ 0.65}_{- 0.88}$ & $-1.45^{+ 0.62}_{- 0.84}$ &sp&sf& r \\
6057  &  2005if  & $03^h30^m12.888^s$ & $-00^{\circ}58^`28.17^{``}$ & $ 9.99^{+ 0.20}_{- 0.01}$ & $ 0.81^{+ 0.02}_{- 0.32}$ &sp&sf&   \\
6304  &  2005jk  & $01^h45^m59.736^s$ & $+01^{\circ}11^`44.63^{``}$ & $10.80^{+ 0.06}_{- 0.35}$ & $ 1.02^{+ 0.04}_{- 0.38}$ &sp&sf& r \\
6406  &  2005ij  & $03^h04^m21.264^s$ & $-01^{\circ}03^`46.87^{``}$ & $10.22^{+ 0.17}_{- 0.06}$ & $ 0.19^{+ 0.45}_{- 0.07}$ &sp&sf& r \\
6422  &  2005id  & $23^h16^m33.360^s$ & $-00^{\circ}39^`48.13^{``}$ & $ 9.52^{+ 0.17}_{- 0.26}$ & $ 0.61^{+ 0.08}_{- 0.59}$ &sp&sf& r \\
6780  &  2005iz  & $21^h52^m16.560^s$ & $+00^{\circ}16^`01.48^{``}$ & $ 8.24^{+ 0.23}_{- 0.24}$ & $-0.32^{+ 0.08}_{- 0.67}$ &sp&sf& r \\
6936  &  2005jl  & $21^h32^m56.160^s$ & $-00^{\circ}42^`00.21^{``}$ & $ 9.89^{+ 0.01}_{- 0.01}$ & $ 0.08^{+ 0.01}_{- 0.01}$ &sp&sf& r \\
7243  &  2005jm  & $21^h52^m18.960^s$ & $+00^{\circ}28^`18.90^{``}$ & $ 8.99^{+ 0.17}_{- 0.11}$ & $ 0.36^{+ 0.07}_{- 0.65}$ &sp&sf& r \\
7876  &  2005ir  & $01^h16^m43.752^s$ & $+00^{\circ}47^`40.36^{``}$ & $ 8.40^{+ 0.30}_{- 0.53}$ & $-0.85^{+ 0.61}_{- 0.16}$ &sp&sf& r \\
8719  &  2005kp  & $00^h30^m53.153^s$ & $-00^{\circ}43^`07.77^{``}$ & $ 8.59^{+ 0.16}_{- 0.07}$ & $-0.04^{+ 0.07}_{- 0.28}$ &sp&sf& r \\
8921  &  2005ld  & $21^h40^m00.480^s$ & $-00^{\circ}00^`28.96^{``}$ & $10.07^{+ 0.22}_{- 0.17}$ & $ 0.78^{+ 0.09}_{- 0.52}$ &sp&sf& r \\
12853  &  2006ey  & $21^h07^m03.600^s$ & $+00^{\circ}43^`24.28^{``}$ & $10.52^{+ 0.06}_{- 0.12}$ & $ 1.00^{+ 0.07}_{- 0.17}$ &sp&sf& r \\
12856  &  2006fl  & $22^h11^m27.600^s$ & $+00^{\circ}45^`20.16^{``}$ & $10.29^{+ 0.22}_{- 0.01}$ & $ 1.00^{+ 0.09}_{- 0.14}$ &sp&sf& r \\
12860  &  2006fc  & $21^h34^m46.800^s$ & $+01^{\circ}10^`31.51^{``}$ & $10.41^{+ 0.23}_{- 0.05}$ & $ 0.13^{+ 0.32}_{- 0.33}$ &sp&sf&   \\
12898  &  2006fw  & $01^h47^m10.344^s$ & $-00^{\circ}08^`48.71^{``}$ & $10.04^{+ 0.10}_{- 0.29}$ & $ 0.27^{+ 0.05}_{- 0.39}$ &sp&sf& r \\
12930  &  2006ex  & $20^h38^m43.920^s$ & $-00^{\circ}28^`34.98^{``}$ & $11.04^{+ 0.07}_{- 0.47}$ & $ 1.39^{+ 0.03}_{- 0.71}$ &sp&sf&   \\
12950  &  2006fy  & $23^h26^m40.080^s$ & $-00^{\circ}50^`26.16^{``}$ & $ 9.82^{+ 0.18}_{- 0.12}$ & $ 0.66^{+ 0.18}_{- 0.37}$ &sp&sf& r \\
13044  &  2006fm  & $22^h10^m10.320^s$ & $+00^{\circ}30^`14.12^{``}$ & $ 9.67^{+ 0.18}_{- 0.01}$ & $ 0.49^{+ 0.02}_{- 0.32}$ &sp&sf& r \\
13070  &  2006fu  & $23^h51^m08.400^s$ & $-00^{\circ}44^`47.64^{``}$ & $10.19^{+ 0.26}_{- 0.01}$ & $ 0.90^{+ 0.09}_{- 0.15}$ &sp&sf&   \\
13152  &  2006gg  & $00^h28^m12.514^s$ & $+00^{\circ}07^`04.77^{``}$ & $ 9.27^{+ 0.00}_{- 0.10}$ & $-0.23^{+ 0.44}_{- 0.01}$ &sp&sf& r \\
13354  &  2006hr  & $01^h50^m15.528^s$ & $-00^{\circ}53^`12.09^{``}$ & $10.69^{+ 0.11}_{- 0.45}$ & $ 1.01^{+ 0.04}_{- 0.74}$ &sp&sf& r \\
13411  &  N/A  & $21^h00^m45.600^s$ & $+00^{\circ}11^`30.17^{``}$ & $ 9.09^{+ 0.23}_{- 0.16}$ & $-0.20^{+ 0.09}_{- 0.51}$ &sp&sf&   \\
13736  &  2006hv  & $22^h27^m19.920^s$ & $+01^{\circ}01^`50.59^{``}$ & $ 9.50^{+ 0.22}_{- 0.01}$ & $ 0.21^{+ 0.09}_{- 0.14}$ &sp&sf& r \\
13796  &  2006hl  & $23^h22^m46.080^s$ & $+00^{\circ}31^`56.34^{``}$ & $10.12^{+ 0.34}_{- 0.01}$ & $ 0.63^{+ 0.20}_{- 0.01}$ &sp&sf& r \\
13894  &  2006jh  & $00^h06^m45.742^s$ & $-00^{\circ}02^`12.29^{``}$ & $ 9.27^{+ 0.34}_{- 0.23}$ & $-0.60^{+ 0.43}_{- 0.40}$ &sp&sf&   \\
14108  &  2006hu  & $03^h34^m22.728^s$ & $-01^{\circ}07^`23.30^{``}$ & $ 8.52^{+ 0.16}_{- 0.06}$ & $-1.28^{+ 0.57}_{- 0.05}$ &sp&sf& r \\
14212  &  2006iy  & $22^h01^m53.040^s$ & $+01^{\circ}02^`40.16^{``}$ & $10.22^{+ 0.05}_{- 0.15}$ & $-0.06^{+ 0.09}_{- 0.44}$ &sp&sf& r \\
14871  &  2006jq  & $03^h37^m06.456^s$ & $+00^{\circ}00^`33.38^{``}$ & $ 9.27^{+ 0.16}_{- 0.01}$ & $ 0.10^{+ 0.02}_{- 0.32}$ &sp&sf& r \\
14979  &  2006jr  & $03^h39^m47.160^s$ & $+00^{\circ}59^`31.68^{``}$ & $ 9.97^{+ 0.26}_{- 0.01}$ & $ 0.68^{+ 0.09}_{- 0.14}$ &sp&sf& r \\
15129  &  2006kq  & $21^h15^m36.480^s$ & $-00^{\circ}19^`18.13^{``}$ & $10.80^{+ 0.33}_{- 0.13}$ & $ 0.78^{+ 0.41}_{- 0.29}$ &sp&sf& r \\
15132  &  2006jt  & $21^h58^m48.240^s$ & $+00^{\circ}11^`54.65^{``}$ & $ 7.40^{+ 0.48}_{- 0.74}$ & $-1.42^{+ 0.65}_{- 0.56}$ &sp&sf& r \\
15136  &  2006ju  & $23^h24^m38.880^s$ & $-00^{\circ}43^`04.59^{``}$ & $11.19^{+ 0.04}_{- 0.06}$ & $ 1.51^{+ 0.02}_{- 0.11}$ &sp&sf& r \\
15234  &  2006kd  & $01^h07^m49.944^s$ & $+00^{\circ}49^`42.89^{``}$ & $10.32^{+ 0.35}_{- 0.16}$ & $ 0.63^{+ 0.30}_{- 0.45}$ &sp&sf&   \\
15259  &  2006kc  & $22^h30^m10.560^s$ & $-00^{\circ}24^`28.04^{``}$ & $ 8.89^{+ 0.45}_{- 0.07}$ & $-1.00^{+ 0.61}_{- 0.04}$ &sp&sf& r \\
15421  &  2006kw  & $02^h14^m57.912^s$ & $+00^{\circ}36^`09.80^{``}$ & $10.17^{+ 0.12}_{- 0.10}$ & $ 0.80^{+ 0.18}_{- 0.22}$ &sp&sf& r \\
15443  &  2006lb  & $03^h19^m28.176^s$ & $-00^{\circ}19^`04.77^{``}$ & $10.48^{+ 0.04}_{- 0.27}$ & $ 0.89^{+ 0.04}_{- 0.16}$ &sp&sf& r \\
15453  &  2006ky  & $21^h18^m40.320^s$ & $-01^{\circ}01^`27.26^{``}$ & $ 8.91^{+ 0.18}_{- 0.19}$ & $-0.27^{+ 0.01}_{- 0.38}$ &sp&sf& r \\
15459  &  2006la  & $22^h42^m48.240^s$ & $-00^{\circ}54^`06.33^{``}$ & $ 8.93^{+ 0.01}_{- 0.01}$ & $ 0.12^{+ 0.01}_{- 0.01}$ &sp&sf& r \\
15461  &  2006kz  & $21^h47^m23.520^s$ & $-00^{\circ}29^`41.15^{``}$ & $10.17^{+ 0.17}_{- 0.05}$ & $ 0.14^{+ 0.33}_{- 0.01}$ &sp&sf& r \\
15467  &    N/A & $21^h20^m04.800^s$ & $-00^{\circ}10^`38.48^{``}$ & $10.40^{+ 0.16}_{- 0.01}$ & $ 1.22^{+ 0.01}_{- 0.26}$ &sp&sf& r \\
15508  &  2006ls  & $01^h48^m40.680^s$ & $-00^{\circ}34^`32.70^{``}$ & $ 9.86^{+ 0.19}_{- 0.01}$ & $ 0.68^{+ 0.02}_{- 0.32}$ &sp&sf& r \\
15583  &  2006mv  & $02^h30^m55.464^s$ & $+00^{\circ}56^`46.62^{``}$ & $ 9.06^{+ 0.30}_{- 0.23}$ & $-0.23^{+ 0.10}_{- 0.52}$ &sp&sf& r \\
15872  &  2006nb  & $02^h26^m53.376^s$ & $-00^{\circ}19^`40.24^{``}$ & $ 9.45^{+ 0.01}_{- 0.04}$ & $-0.04^{+ 0.06}_{- 0.01}$ &sp&sf& r \\
16021  &  2006nc  & $00^h55^m22.512^s$ & $-00^{\circ}23^`21.16^{``}$ & $ 9.84^{+ 0.47}_{- 0.06}$ & $-0.12^{+ 0.74}_{- 0.02}$ &sp&sf& r \\
16069  &  2006nd  & $22^h44^m58.800^s$ & $-01^{\circ}00^`22.97^{``}$ & $11.09^{+ 0.02}_{- 0.45}$ & $ 1.14^{+ 0.09}_{- 1.00}$ &sp&sf&   \\
16073  &  2006of  & $00^h32^m25.838^s$ & $-01^{\circ}03^`14.04^{``}$ & $ 9.78^{+ 0.05}_{- 0.18}$ & $ 0.41^{+ 0.01}_{- 0.23}$ &sp&sf& r \\
16099  &  2006nn  & $01^h45^m41.112^s$ & $-01^{\circ}03^`16.45^{``}$ & $10.44^{+ 0.28}_{- 0.10}$ & $ 0.28^{+ 0.50}_{- 0.51}$ &sp&sf&   \\
16116  &  2006oh  & $23^h41^m12.480^s$ & $-01^{\circ}06^`21.46^{``}$ & $ 9.41^{+ 0.18}_{- 0.01}$ & $ 0.23^{+ 0.02}_{- 0.33}$ &sp&sf& r \\
16276  &  2006om  & $01^h22^m18.840^s$ & $+01^{\circ}00^`37.26^{``}$ & $ 9.32^{+ 0.16}_{- 0.37}$ & $-0.25^{+ 0.24}_{- 0.38}$ &sp&sf& r \\
16287  &  2006np  & $03^h06^m39.576^s$ & $+00^{\circ}03^`43.15^{``}$ & $11.22^{+ 0.16}_{- 0.32}$ & $ 1.54^{+ 0.11}_{- 0.25}$ &sp&sf& r \\
16314  &  2006oa  & $21^h23^m42.960^s$ & $-00^{\circ}50^`35.02^{``}$ & $ 9.01^{+ 0.07}_{- 0.04}$ & $-0.44^{+ 0.30}_{- 0.07}$ &sp&sf&   \\
16618  &  2006pq  & $01^h25^m06.888^s$ & $-01^{\circ}13^`09.80^{``}$ & $10.10^{+ 0.05}_{- 0.17}$ & $ 0.81^{+ 0.01}_{- 0.52}$ &sp&sf& r \\
16619  &  2006ps  & $01^h43^m45.312^s$ & $-01^{\circ}06^`42.62^{``}$ & $ 7.79^{+ 0.12}_{- 0.15}$ & $-1.66^{+ 0.46}_{- 0.04}$ &sp&sf& r \\
17168  &  2007ik  & $22^h38^m53.760^s$ & $-01^{\circ}10^`02.14^{``}$ & $ 9.52^{+ 0.02}_{- 0.01}$ & $ 0.47^{+ 0.01}_{- 0.44}$ &sp&sf& r \\
17186  &  2007hx  & $02^h06^m27.288^s$ & $-00^{\circ}53^`57.57^{``}$ & $ 8.25^{+ 0.33}_{- 0.72}$ & $-0.42^{+ 0.36}_{- 0.70}$ &sp&sf& r \\
17218  &  2007jp  & $23^h44^m41.280^s$ & $-00^{\circ}01^`48.27^{``}$ & $10.02^{+ 0.34}_{- 0.01}$ & $ 0.53^{+ 0.20}_{- 0.01}$ &sp&sf& r \\
17389  &  2007ih  & $21^h33^m10.800^s$ & $-00^{\circ}57^`36.30^{``}$ & $ 9.55^{+ 0.21}_{- 0.01}$ & $-0.09^{+ 0.46}_{- 0.01}$ &sp&sf& r \\
17497  &  2007jt  & $02^h28^m32.760^s$ & $-01^{\circ}02^`34.12^{``}$ & $10.28^{+ 0.30}_{- 0.01}$ & $ 0.89^{+ 0.15}_{- 0.07}$ &sp&sf& r \\
17568  &  2007kb  & $20^h52^m24.720^s$ & $+00^{\circ}16^`38.90^{``}$ & $10.10^{+ 0.07}_{- 0.26}$ & $ 0.45^{+ 0.03}_{- 0.29}$ &sp&sf&   \\
17745  &  2007ju  & $00^h11^m50.465^s$ & $-00^{\circ}20^`21.67^{``}$ & $ 8.82^{+ 0.05}_{- 0.16}$ & $-0.01^{+ 0.04}_{- 0.29}$ &sp&sf& r \\
17880  &  2007jd  & $02^h59^m53.664^s$ & $+01^{\circ}09^`36.25^{``}$ & $10.39^{+ 0.03}_{- 0.23}$ & $ 0.71^{+ 0.08}_{- 0.24}$ &sp&sf&   \\
18030  &  2007kq  & $00^h19^m43.970^s$ & $-00^{\circ}24^`00.35^{``}$ & $ 9.70^{+ 0.07}_{- 0.11}$ & $ 0.79^{+ 0.04}_{- 0.55}$ &sp&sf& r \\
18241  &  2007ks  & $20^h49^m33.120^s$ & $-00^{\circ}45^`42.94^{``}$ & $ 9.39^{+ 0.23}_{- 0.01}$ & $-0.79^{+ 0.49}_{- 0.30}$ &sp&sf& r \\
18323  &  2007kx  & $00^h13^m42.874^s$ & $+00^{\circ}39^`08.38^{``}$ & $ 9.32^{+ 0.22}_{- 0.05}$ & $ 0.04^{+ 0.09}_{- 0.15}$ &sp&sf& r \\
18602  &  2007lo  & $22^h35^m56.160^s$ & $+00^{\circ}36^`32.79^{``}$ & $ 9.32^{+ 0.01}_{- 0.02}$ & $-0.08^{+ 0.02}_{- 0.15}$ &sp&sf& r \\
18650  &  2007lt  & $21^h53^m47.280^s$ & $+00^{\circ}00^`54.10^{``}$ & $ 8.90^{+ 0.08}_{- 0.01}$ & $-0.15^{+ 0.08}_{- 0.01}$ &sp&sf& r \\
18697  &  2007ma  & $00^h44^m53.808^s$ & $-00^{\circ}59^`48.71^{``}$ & $10.18^{+ 0.38}_{- 0.08}$ & $ 0.31^{+ 0.43}_{- 0.70}$ &sp&sf& r \\
18804  &  2007me  & $01^h41^m03.840^s$ & $-00^{\circ}26^`53.78^{``}$ & $10.25^{+ 0.34}_{- 0.19}$ & $ 0.77^{+ 0.20}_{- 0.51}$ &sp&sf& r \\
18903  &  2007lr  & $00^h49^m00.288^s$ & $-00^{\circ}19^`23.80^{``}$ & $10.92^{+ 0.23}_{- 0.15}$ & $ 0.64^{+ 0.57}_{- 0.44}$ &sp&sf&   \\
19149  &  2007ni  & $02^h05^m50.496^s$ & $-00^{\circ}19^`57.27^{``}$ & $ 9.57^{+ 0.02}_{- 0.05}$ & $ 0.41^{+ 0.08}_{- 0.37}$ &sp&sf& r \\
19543  &  2007oj  & $23^h51^m37.920^s$ & $+00^{\circ}16^`47.38^{``}$ & $ 8.83^{+ 0.14}_{- 0.11}$ & $-0.41^{+ 0.01}_{- 0.34}$ &sp&sf& r \\
19616  &  2007ok  & $02^h28^m23.904^s$ & $+00^{\circ}11^`09.65^{``}$ & $11.16^{+ 0.06}_{- 0.51}$ & $ 1.30^{+ 0.05}_{- 0.74}$ &sp&sf& r \\
19626  &  2007ou  & $02^h23^m42.648^s$ & $-00^{\circ}49^`35.28^{``}$ & $10.45^{+ 0.08}_{- 0.27}$ & $ 0.87^{+ 0.05}_{- 0.17}$ &sp&sf&   \\
19658  &  2007ot  & $00^h35^m36.775^s$ & $-00^{\circ}13^`57.36^{``}$ & $ 8.53^{+ 0.20}_{- 0.02}$ & $-0.28^{+ 0.10}_{- 0.16}$ &sp&sf& r \\
19899  &  2007pu  & $22^h45^m58.320^s$ & $-00^{\circ}38^`55.99^{``}$ & $ 8.68^{+ 0.05}_{- 0.17}$ & $-0.15^{+ 0.07}_{- 0.61}$ &sp&sf& r \\
19913  &  2007qf  & $22^h15^m02.880^s$ & $-00^{\circ}20^`30.23^{``}$ & $ 9.89^{+ 0.05}_{- 0.01}$ & $ 0.60^{+ 0.24}_{- 0.20}$ &sp&sf& r \\
19940  &  2007pa  & $21^h01^m34.560^s$ & $-00^{\circ}16^`07.56^{``}$ & $ 8.52^{+ 0.58}_{- 0.09}$ & $-1.12^{+ 0.62}_{- 0.17}$ &sp&sf& r \\
19953  &  2007pf  & $22^h11^m43.200^s$ & $+00^{\circ}34^`45.68^{``}$ & $ 9.65^{+ 0.06}_{- 0.06}$ & $ 0.47^{+ 0.08}_{- 0.36}$ &sp&sf& r \\
19969  &  2007pt  & $02^h07^m38.352^s$ & $-00^{\circ}19^`26.50^{``}$ & $10.50^{+ 0.10}_{- 0.18}$ & $ 1.13^{+ 0.01}_{- 0.23}$ &sp&sf& r \\
20084  &  2007pd  & $23^h11^m54.240^s$ & $-00^{\circ}34^`44.61^{``}$ & $10.18^{+ 0.18}_{- 0.13}$ & $ 0.93^{+ 0.02}_{- 0.34}$ &sp&sf&   \\
20350  &  2007ph  & $20^h51^m13.200^s$ & $-00^{\circ}57^`17.86^{``}$ & $10.98^{+ 0.01}_{- 0.01}$ & $ 2.35^{+ 0.01}_{- 0.01}$ &sp&sf&   \\
20430  &  2007qj  & $20^h49^m40.080^s$ & $+00^{\circ}28^`06.65^{``}$ & $ 8.08^{+ 0.38}_{- 0.12}$ & $-1.12^{+ 0.44}_{- 0.20}$ &sp&sf& r \\
20764  &  2007ro  & $01^h44^m28.992^s$ & $+00^{\circ}13^`47.26^{``}$ & $10.83^{+ 0.26}_{- 0.18}$ & $ 0.67^{+ 0.24}_{- 0.51}$ &sp&sf& r \\
20821  &  2007rk  & $03^h42^m17.376^s$ & $+01^{\circ}03^`44.17^{``}$ & $10.64^{+ 0.01}_{- 0.26}$ & $ 0.46^{+ 0.01}_{- 0.32}$ &sp&sf&   \\
21062  &  2007rp  & $22^h13^m43.680^s$ & $+00^{\circ}23^`46.69^{``}$ & $ 9.27^{+ 0.08}_{- 0.12}$ & $-0.20^{+ 0.21}_{- 0.14}$ &sp&sf& r \\
21810  &  2007se  & $22^h12^m37.200^s$ & $+00^{\circ}47^`49.37^{``}$ & $ 8.70^{+ 0.02}_{- 0.15}$ & $-0.36^{+ 0.01}_{- 0.41}$ &sp&sf&   \\
1740  & N/A & $00^h21^m37.054^s$ & $-00^{\circ}52^`51.30^{``}$ & $10.71^{+ 0.21}_{- 0.10}$ & N/A &lc& p& r \\
2162  & N/A & $01^h01^m46.296^s$ & $-00^{\circ}08^`01.43^{``}$ & $10.93^{+ 0.19}_{- 0.12}$ & N/A &lc& p& r \\
3488  & N/A & $20^h54^m13.200^s$ & $-01^{\circ}00^`37.26^{``}$ & $ 9.83^{+ 0.06}_{- 0.01}$ & N/A &lc& p&   \\
4065  & N/A & $03^h11^m59.544^s$ & $+01^{\circ}03^`04.57^{``}$ & $10.81^{+ 0.06}_{- 0.01}$ & N/A &lc& p& r \\
4690  & N/A & $02^h11^m43.104^s$ & $+00^{\circ}41^`17.66^{``}$ & $10.36^{+ 0.01}_{- 0.01}$ & N/A &lc& p&   \\
5785  & N/A & $21^h54^m23.520^s$ & $+00^{\circ}05^`03.70^{``}$ & $11.68^{+ 0.15}_{- 0.12}$ & N/A &lc& p&   \\
6530  & N/A & $00^h57^m18.984^s$ & $+00^{\circ}01^`16.88^{``}$ & $ 9.89^{+ 0.01}_{- 0.01}$ & N/A &lc& p& r \\
7600  & N/A & $23^h10^m27.360^s$ & $-01^{\circ}06^`27.68^{``}$ & $10.81^{+ 0.25}_{- 0.10}$ & N/A &lc& p& r \\
13907  & N/A & $00^h56^m43.080^s$ & $+00^{\circ}13^`57.48^{``}$ & $10.65^{+ 0.38}_{- 0.02}$ & N/A &lc& p&   \\
15748  & N/A & $03^h12^m27.504^s$ & $-00^{\circ}07^`50.33^{``}$ & $10.70^{+ 0.07}_{- 0.01}$ & N/A &lc& p& r \\
15866  & N/A & $22^h35^m07.440^s$ & $+00^{\circ}58^`05.18^{``}$ & $10.98^{+ 0.10}_{- 0.07}$ & N/A &lc& p& r \\
16103  & N/A & $20^h51^m54.000^s$ & $-01^{\circ}03^`00.36^{``}$ & $10.09^{+ 0.16}_{- 0.02}$ & N/A &lc& p& r \\
17928  & N/A & $23^h53^m52.800^s$ & $+01^{\circ}06^`51.30^{``}$ & $10.94^{+ 0.38}_{- 0.02}$ & N/A &lc& p&   \\
911  & N/A & $02^h34^m45.840^s$ & $-00^{\circ}06^`54.97^{``}$ & $10.29^{+ 0.07}_{- 0.29}$ & $ 0.80^{+ 0.01}_{- 0.21}$ &lc&sf&   \\
3049  & N/A & $22^h00^m53.520^s$ & $-01^{\circ}14^`11.58^{``}$ & $ 9.89^{+ 0.01}_{- 0.01}$ & $ 0.25^{+ 0.01}_{- 0.01}$ &lc&sf& r \\
3959  & N/A & $03^h00^m23.472^s$ & $+00^{\circ}30^`43.06^{``}$ & $ 9.50^{+ 0.01}_{- 0.21}$ & $ 0.11^{+ 0.01}_{- 0.46}$ &lc&sf&   \\
4019  & N/A & $00^h05^m02.854^s$ & $+01^{\circ}08^`47.08^{``}$ & $10.86^{+ 0.34}_{- 0.17}$ & $ 0.99^{+ 0.35}_{- 0.33}$ &lc&sf&   \\
5689  & N/A & $23^h51^m31.920^s$ & $+00^{\circ}49^`39.16^{``}$ & $10.31^{+ 0.35}_{- 0.16}$ & $ 0.62^{+ 0.30}_{- 0.45}$ &lc&sf&   \\
5959  & N/A & $02^h32^m14.304^s$ & $-00^{\circ}18^`28.96^{``}$ & $10.88^{+ 0.01}_{- 0.01}$ & $ 0.60^{+ 0.01}_{- 0.01}$ &lc&sf& r \\
6614  & N/A & $01^h46^m35.304^s$ & $+00^{\circ}52^`01.86^{``}$ & $10.67^{+ 0.32}_{- 0.01}$ & $ 0.98^{+ 0.30}_{- 0.01}$ &lc&sf&   \\
6861  & N/A & $23^h17^m43.680^s$ & $-01^{\circ}06^`49.39^{``}$ & $ 9.92^{+ 0.21}_{- 0.02}$ & $-0.27^{+ 0.39}_{- 0.33}$ &lc&sf&   \\
8254  & N/A & $23^h24^m39.360^s$ & $+00^{\circ}49^`11.90^{``}$ & $ 9.63^{+ 0.11}_{- 0.01}$ & $ 0.58^{+ 0.01}_{- 0.44}$ &lc&sf& r \\
8280  & N/A & $00^h34^m17.558^s$ & $+00^{\circ}47^`44.02^{``}$ & $10.48^{+ 0.34}_{- 0.01}$ & $ 0.99^{+ 0.20}_{- 0.01}$ &lc&sf&   \\
8555  & N/A & $00^h11^m39.744^s$ & $-00^{\circ}24^`54.12^{``}$ & $ 9.96^{+ 0.09}_{- 0.10}$ & $ 0.59^{+ 0.12}_{- 0.22}$ &lc&sf&   \\
11172  & N/A & $21^h29^m39.120^s$ & $-00^{\circ}12^`07.88^{``}$ & $10.18^{+ 0.14}_{- 0.03}$ & $ 1.00^{+ 0.02}_{- 0.33}$ &lc&sf& r \\
11311  & N/A & $03^h08^m03.696^s$ & $+00^{\circ}26^`00.51^{``}$ & $10.58^{+ 0.33}_{- 0.13}$ & $ 0.56^{+ 0.41}_{- 0.29}$ &lc&sf& r \\
12804  & N/A & $01^h12^m48.408^s$ & $+01^{\circ}02^`24.61^{``}$ & $ 9.48^{+ 0.11}_{- 0.01}$ & $ 0.30^{+ 0.02}_{- 0.32}$ &lc&sf& r \\
14317  & N/A & $21^h02^m17.040^s$ & $+00^{\circ}19^`50.15^{``}$ & $10.77^{+ 0.01}_{- 0.01}$ & $ 0.49^{+ 0.01}_{- 0.01}$ &lc&sf& r \\
14525  & N/A & $01^h07^m31.872^s$ & $+00^{\circ}28^`38.80^{``}$ & $10.16^{+ 0.08}_{- 0.27}$ & $ 0.58^{+ 0.05}_{- 0.17}$ &lc&sf&   \\
14784  & N/A & $21^h35^m11.760^s$ & $-00^{\circ}20^`55.90^{``}$ & $10.34^{+ 0.13}_{- 0.01}$ & $ 0.15^{+ 0.39}_{- 0.01}$ &lc&sf& r \\
15892  & N/A & $21^h32^m47.760^s$ & $+00^{\circ}41^`19.94^{``}$ & $10.95^{+ 0.07}_{- 0.45}$ & $ 1.01^{+ 0.11}_{- 0.69}$ &lc&sf&   \\
16163  & N/A & $02^h05^m59.808^s$ & $-00^{\circ}51^`20.97^{``}$ & $10.50^{+ 0.01}_{- 0.17}$ & $ 0.63^{+ 0.01}_{- 0.39}$ &lc&sf&   \\
17434  & N/A & $01^h13^m45.048^s$ & $-00^{\circ}04^`22.60^{``}$ & $10.51^{+ 0.06}_{- 0.19}$ & $ 1.03^{+ 0.04}_{- 0.16}$ &lc&sf& r \\
17748  & N/A & $00^h39^m13.673^s$ & $-00^{\circ}16^`39.05^{``}$ & $ 9.89^{+ 0.11}_{- 0.01}$ & $ 0.71^{+ 0.01}_{- 0.32}$ &lc&sf&   \\
20545  & N/A & $22^h01^m00.000^s$ & $-00^{\circ}25^`26.35^{``}$ & $10.22^{+ 0.51}_{- 0.06}$ & $ 0.19^{+ 0.71}_{- 0.01}$ &lc&sf   \\
\enddata
\end{deluxetable}

\end{document}